\documentclass[pre,twocolumn,preprintnumbers,showpacs,floatfix,superscriptaddress]{revtex4-1}
\usepackage{epsf}
\usepackage{graphicx}
\usepackage{dcolumn}
\usepackage{bm}
\usepackage{subfigure}
\usepackage{color}
\usepackage{amsthm}
\usepackage{amssymb}
\usepackage{amsmath}
\usepackage{pb-diagram}

\usepackage{psfig}
\usepackage{dcolumn}
\usepackage{bm}
\usepackage{subfigure}
\usepackage{color}
\usepackage{amsthm}
\usepackage{amssymb}
\usepackage{amsmath}
\usepackage{pb-diagram}

\begin{document}

\preprint{}
\title[Force networks in disks and pentagons]{Structure of force networks in tapped particulate systems of disks and pentagons (Part 1): Clusters and Loops}

\author{Luis A. Pugnaloni}
\affiliation{Dpto.~de Ingenier\'ia Mec\'anica, Facultad Regional La Plata, Universidad Tecnol\'ogica Nacional, Av. 60 Esq. 124, 1900 La Plata, Argentina}
\affiliation{Consejo Nacional de Investigaciones Cient\'ificas y T\'ecnicas, Argentina}

\author{C. Manuel Carlevaro}
\affiliation{Instituto de F\'isica de L\'iquidos y Sistemas Biol\'ogicos (CONICET La Plata,
UNLP), Calle 59 Nro 789, 1900 La Plata, Argentina}
\affiliation{Universidad Tecnol\'ógica Nacional-FRBA, UDB F\'isica, Mozart 2300, C1407IVT Buenos Aires, Argentina}

\author{M. Kram\'ar}
\author{K. Mischaikow}
\affiliation{Department of Mathematics, Rutgers University, Piscataway, New Jersey 08854-8019, USA}

\author{L. Kondic}
\affiliation{Department of Mathematical Sciences, New Jersey Institute of Technology, Newark, New Jersey 07102, USA}

\begin{abstract}
The force network of a granular assembly, defined by the contact network and the corresponding contact forces, 
carries valuable information about the state of the packing. Simple analysis of these networks based on the distribution of force strengths 
is rather insensitive to the changes in preparation protocols or to the types of particles. 
In this paper we consider two dimensional simulations of tapped systems built from frictional disks and pentagons, and 
study the structure of the force networks of granular packings by considering  network's topology as force thresholds are 
varied.  We show that the number of clusters and loops observed in the force networks as a function of the force threshold are 
markedly different for disks and pentagons if the tangential contact forces are considered, whereas they are surprisingly similar for the network
defined by the normal forces.  In particular, the  results indicate that, overall, the force network is more heterogeneous for disks than for pentagons. 
Additionally, we show that the states obtained  by tapping with different intensities that display similar packing fraction are difficult to distinguish based 
on simple topological invariants.
\end{abstract}

\volumeyear{}
\volumenumber{}
\issuenumber{}
\eid{}
\date{\today}
\startpage{1}
\endpage{}
\maketitle

\section{Introduction}

Particulate systems are very common in nature and in a variety of technologically
relevant applications.     Many of these systems are composed of particles that remain in contact for relatively long periods.   These contacts form 
a network, whose properties are important for the purpose of understanding 
the system as a whole.  However, the contact network  provides only partial information about the interaction between the particles. In order to obtain a deeper understanding of a particulate system, the  strength of the contacts needs to be considered. This naturally leads to the concept of force networks.  The properties of the force networks 
are of fundamental importance for the purpose  of revealing the underlying physical causes of many phenomena. For example, the electrical conductivity of a granular bed strongly depends not only on the actual strength of the contacts \cite{falcon2004nonlinear}, but also on the structure of the contact network and the presence of paths of strong contacts \cite{dorbolo2002electrical}. Similarly, the elastic properties of these systems are very sensitive to the characteristic of underlying force network \cite{jia_prl99,makse_pre04}, to the degree that different packing structures can be distinguished by their response to sound propagation \cite{lherminier2014revealing}.

Due to their importance, both contact and force networks have been analyzed
extensively in recent years. While earlier research focused mostly on basic statistical properties of these
networks by computing probability density functions (PDFs) of the contact forces, see, e.g.,~\cite{radjai_96b, majmudar05a}, during the last few years a
variety of other approaches have been considered.   These approaches include, among others, network-type of analysis~\cite{daniels_pre12, herrera_pre11,walker_pre12,bassett_soft_mat15}, exploration of the properties 
of the cycles (loops) formed by the force~\cite{tordesillas_pre10} and contact~\cite{arevalo_pre10, arevalo_pre13} networks,
and the force tiling approach~\cite{tighe_jsm11}.    Regarding
the connection (or the lack of it) between force and contact networks,  it is worth
mentioning recent works~\cite{zhang_softmatter13,carlevaro2012arches} that illustrated that the properties of 
`force chains' (defined appropriately) differ significantly from the dominant geometrical
features (bridges/arches) arising in the contact network.    The analysis of topological properties
of force and contact networks~\cite{epl12,pre13,pre14,physicaD14} has quantified the differences between these networks 
in much more detail, and has generally shown that the properties of the force networks
depend strongly on the force level (threshold) considered. 

Particulate systems consisting of circular/spherical particles have been intensively studied by theoretical, computational  and experimental approaches. 
The simple shape of these particles makes the study easier compared to the systems consisting of particles of other shapes. However, the particles 
relevant to applications are typically not circular.  Therefore, it is important to understand similarities, as well as differences, between the 
systems consisting of  circular/spherical particles and particles of other shapes. This problem appears to be  rather complex, and the number of
works considering in detail the influence of the particle shape is rather limited, with focus on probability density functions (PDFs) of the 
inter-particle forces~\cite{radjai_09,radjai_pre14},  contact networks~\cite{azema_pre13}, anisotropy of force networks 
under shear~\cite{hidalgo2009role,hidalgo2010granular,azema2007force}, and influence of particle shape on 
shear strength of the material~\cite{cegeo_epl2012}.  
Connectivity properties were also discussed in the context of experiments carried out with photoelastic particles~\cite{behringer_2001}. 
Considering the force PDF for a moment, we note that this measure only provides information about the  strength of the 
forces and does not shed much light on the local and global properties of the underlying force network. We will show that 
physical systems with  similar PDFs can give rise to very different force networks and therefore could exhibit different response to 
external perturbation.

\begin{figure}[thb]
\centering
\includegraphics[width=1.0\columnwidth]{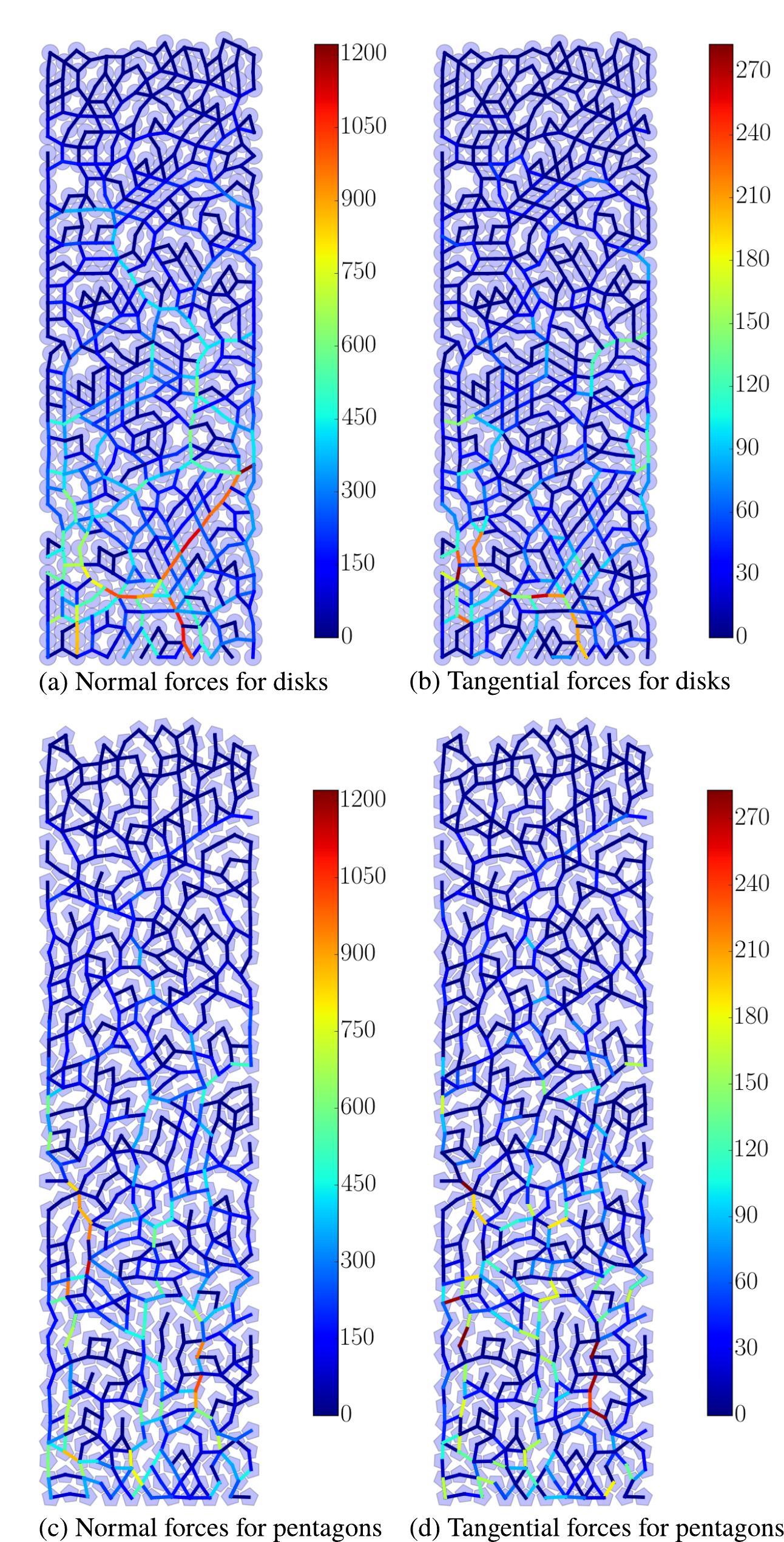}
\caption{(Color online) Sample snapshots of the disks (a) and (b), and pentagons (c) and (d) prepared by tapping at 
low tapping intensity, $\Gamma=3.83\sqrt{dg}$ (see Sect. \ref{sec:methods}). The contact forces are indicated by segments 
connecting the touching particles colored according to the strength of the normal (a) and (c), or tangential (b) and (d) components of the force. 
The color scale indicates the force in units of $mg$, with $m$ the mass of a particle and $g$ the acceleration of gravity.}
\label{fig:snap}
\end{figure}

In this and the companion paper~\cite{paper2}, we focus on quantifying force networks 
in tapped systems of disks and pentagons in two spatial dimensions.  Figure~\ref{fig:snap} shows examples of the force networks that we will consider.
In addition to quantifying the influence of particle shape on the properties of force networks,
we also consider the spatial heterogeneity induced by gravity, as well as the effect of the preparation protocol on seemingly equivalent states.  For this 
purpose, we use the techniques
based on computational topology that have been already used to analyze systems
exposed to compression~\cite{epl12,pre13,pre14,physicaD14}.   In the present paper, we will consider simple
measures, based on analyzing the components (clusters) and the loops (holes) formed
by the force networks at different force thresholds. We focus on the Betti numbers $\beta_0$ and $\beta_1$, that correspond to the number of components and loops, respectively.  Additional more complex measures based on persistence homology are discussed in~\cite{paper2}.  

One interesting feature of granular columns subjected to vibration (or taps) is that they reach steady states that may have the same packing fraction, $\phi$, even if the intensity of the taps differ significantly \cite{pugnaloni2008nonmonotonic,pugnaloni_pre10,pugnaloni2011master}. This is the case also for polygonal grains \cite{carlevaro_jsm11}. Hence, a natural question is what distinguishes two states with same $\phi$. 
Here, we will go beyond some preliminary findings \cite{arevalo_pre13} on the distinctive contact network of these same-$\phi$ steady states and look into the associated force network. 

After presenting the computational methods in Sec.~\ref{sec:methods}, we  compare force networks in different settings. 
In  Sec.~\ref{sec:up_down}, we show  that the gravitational compaction influences the structure of the force networks beyond a simple change in the average force level. 
In Sec.~\ref{sec:high_low} we turn our attention to comparing the force networks 
of a system characterized by the same packing fraction obtained with different tap intensities. Then, in Sec.~\ref{sec:disks_pents},  we identify similarities and differences between  force networks  corresponding to  the systems built out of disks and pentagons.    Section~\ref{sec:conclusions} is devoted to conclusions and future outlook.

\section{Methods}
\label{sec:methods}

\subsection{Simulation}
We carried out discrete element method (DEM) simulations of two-dimensional (2D) systems of particles. The simulations were implemented by means of the Box2D library \cite{box2d}. Box2D uses a constraint solver to handle hard bodies. At each time step of the dynamics a series of iterations (typically 20) are used to resolve overlaps between bodies through a Lagrange multiplier scheme \cite{catto}. After resolving overlaps, the inelastic collision at each contact   is solved and new linear and angular velocities are assigned. The equations of motion are integrated through a symplectic Euler algorithm. Solid friction is also handled by means of a Lagrange multiplier scheme that implements the Coulomb criterion. This library achieves a high performance when handling complex bodies such as polygons. The approach yields realistic dynamics for granular complex bodies and has been successfully used to study grains under tapping \cite{carlevaro_jsm11,
irastorza2013exact} and also under vigorous vibration \cite{sanchez2014effect}.

The systems consist of 500 monosized particles (either disks or regular pentagons). Both type of particles have the same area and material density. The diameter, $d$, of the disks is set to $1.0$. Pentagons have a radius (center-to-vertex distance) of $0.57474d$; this choice ensures that they have the same mass as disks. The particles are placed in a rectangular box $14.4d$ wide which is confined to move in the vertical direction. This box is high enough to avoid particles contacts with the ceiling. 

We set the particle--particle interactions to yield a low normal restitution coefficient $\epsilon=0.058$. This ensures that the grains come to rest after each tap in a relatively short simulation time. The static, $\mu_s$, and dynamic, $\mu_d$, friction coefficients are set to $0.5$. The particle--wall restitution coefficient is as in the particle--particle interaction. The particle--wall friction coefficients are $\mu_s = \mu_d = 0.005$. These low values lead to a reduced Janssen-like effect. We define $m$ as the mass of a particle and $g$ as the acceleration of gravity. The time step $\delta t$ used to integrate the equations of motion is $0.015 \sqrt{d/g}$.

Particles, initially placed at random, without overlaps in the box, are let to settle until they come to rest in order to prepare the initial packing. Then, $600$ taps are applied to each sample. After each tap, we wait for the particles to equilibrate. The new static configuration after each tap is saved for the force network analysis. This includes the particle positions and orientations along with the contact forces. In the case of pentagons, Box2D represents side-to-side contacts as two effective point contacts that define a segment along the shared side. We add these two forces on both point contacts to represent the total force exerted on the face-to-face contact. The initial $100$ taps are discarded and only the final $500$ ones are used in the analysis. In previous works~\cite{pugnaloni_pre10,pugnaloni2011master},
we found that hundred taps was sufficient to reach a well defined steady state for the tap intensities used.

\begin{figure}[]
\centering
\includegraphics[width=0.90\columnwidth]{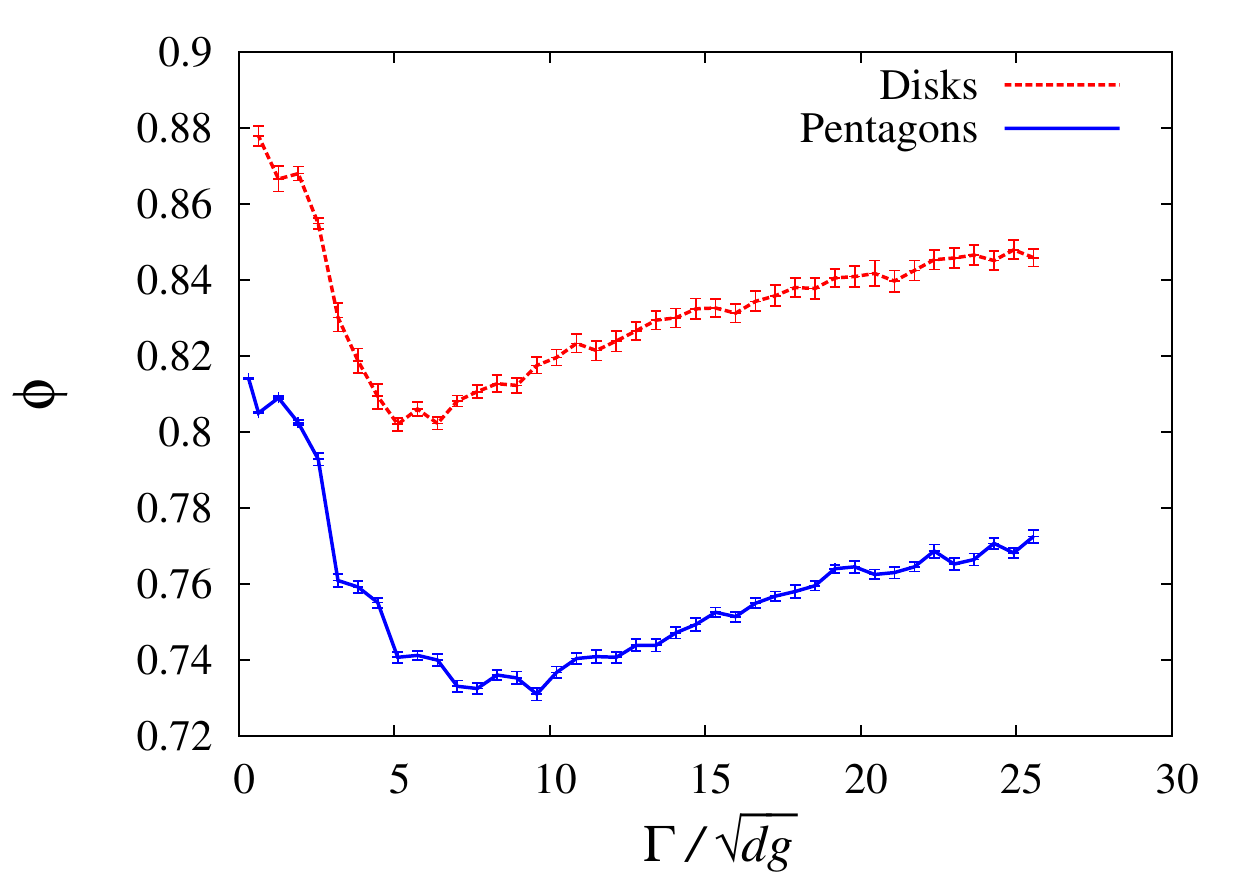}
\caption{(Color online) Steady state packing fraction, $\phi$, as a function of $\Gamma$ for disks and pentagons. The error bars indicate the standard error on the mean $\phi$, which is averaged over $30$ taps after reaching the steady state.}
\label{fig:phi}
\end{figure}

Tapping itself is simulated by hitting the containing box upwards so it flies and falls back due to gravity on a solid base. To achieve this, while the box is at rest on the solid base, we reset the 
velocity, $v_0$, of the box to a given positive value and restart the dynamics.  The box--base restitution coefficient is set to zero. While the box dissipate all its kinetic energy when contacting the base, particles inside the box bounce against the box walls and floor until settling. After all particles come to rest, a new tap is applied. The intensity, $\Gamma$, of the taps is measured by the initial velocity 
imposed to the confining box at each tap (i.e. $\Gamma=v_0$).   

Figure~\ref{fig:phi} shows the steady state $\phi$ as a function of $\Gamma$.   In this paper, we consider two values of $\Gamma$,  
$\Gamma=3.83\sqrt{dg}$ (low tapping intensity), and $\Gamma = 12.14\sqrt{dg}$ (high tapping intensity).   
Although pentagons display much lower packing fractions, both systems reach a minimum in $\phi$, as discussed 
in~\cite{carlevaro_jsm11}. The low tapping intensity is low enough to warrant that for both particle shapes the minimum 
$\phi$ has not been reached, yet large enough to avoid the ergodicity breaking observed at very low tap intensities \cite{paillusson2012probing}. 
Due to ergodicity breaking, at lower tap intensities, two independent realizations of the tapping protocol may lead to different, distinguishable, steady states.

Figure~\ref{fig:snap} shows few snapshots of disks and pentagons (particles and force networks).    Since, based on visual inspection, 
the geometrical arrangement and the forces changes dramatically from tap to tap, all our results are computed by averaging over a large number ($500$) of taps.   
This number of realizations is sufficient to decrease the statistical fluctuations significantly, and will allow us to identify the differences between the considered 
networks that are robust with respect to statistical fluctuations. We consider only the particle--particle contacts in 
our analysis; the contacts with the walls and floor of the container are disregarded.

\subsection{Betti numbers}

The force network is represented by a scalar function, $f$, from the contact network to the real numbers. Values of this function at the edges of the network (i.e., the connecting lines between contacting particles) are given by the magnitude of the normal (tangential) force. The function $f$ is normalized by the average force level in the system. So, an edge in the network with the value $1$ represents a contact between two particles with a force value equal to the average force $\langle f \rangle$.

We are interested in the structural properties of the part of the force network on which $f$ exceeds a given  threshold $F$. We study the  properties of those parts of the network for different values of $F$. For simplicity, we restrict our attention to the number of connected components (just `components' 
or `clusters' in what follows) and the number of loops present in the networks. 
In particular, the function $\beta_0(F)$ measures the number of  components for the part of the force network exceeding the force level $F$.  For $F > 1$ these components can be thought of 
as `force chains'.
Additional insight into the properties of a force network can be obtained by considering the number of loops, $\beta_1(F)$, inside of the part of the network exceeding the force level $F$.    A loop 
in the network is a closed path of the edges connecting centers of the particles. 
If not indicated otherwise, we do not consider the trivial loops (i.e., loops made by three contacting particles); furthermore, all the results
for $\beta_0$ and $\beta_1$ {\it are normalized by the number of particles} in the domain used to define the force network under consideration.

A more detailed description of  Betti numbers can be 
found in~\cite{epl12} while in depth discussion of computational homology in the context of particular matter is given in~\cite{physicaD14}.  General treatment of computational algebraic topology can be found in~\cite{mischaikow, edelsbrunner:harer}.

\section{Results}

\subsection{Heterogeneity of force networks in gravitationally compacted systems}
\label{sec:up_down}

We start by investigating the influence of gravity on the structural properties of the force networks corresponding to the steady states of the 
tapped disk-based systems.  The average contact force, $\langle f \rangle$,  increases with depth as long as Janssen's saturation is not achieved. As shown 
in~\cite{karim2014eliminating},  achieving this saturation  is difficult in 2D. Therefore, the average force as well as the structure of the force network might change with depth. In order to understand the influence of the depth on the network structure, we compare two distinct horizontal slices of the system. The thickness of both slices is $10d$ and the forces in each slice are normalized by  $\langle f \rangle$ inside of the slice. We consider only the particles whose centers are inside the slice and the interactions between these particles. The contacts with other particles or walls are not taken into account. Each slice contains roughly $140$ particles, with some variation from tap to tap and from disk-based to pentagon-based systems.
The top slice is centered at $27d$ from the box floor, which ensures that the upper boundary of this slice is about $4d$ below the free surface.  To minimize the boundary effects, the bottom slice is centered at $7d$ from the bottom of the box. 
Thus, approximately three bottom layers of grains are excluded. 

\begin{figure}[]
\centering
\subfigure[Normal forces.]
{\includegraphics[width=0.49\columnwidth]{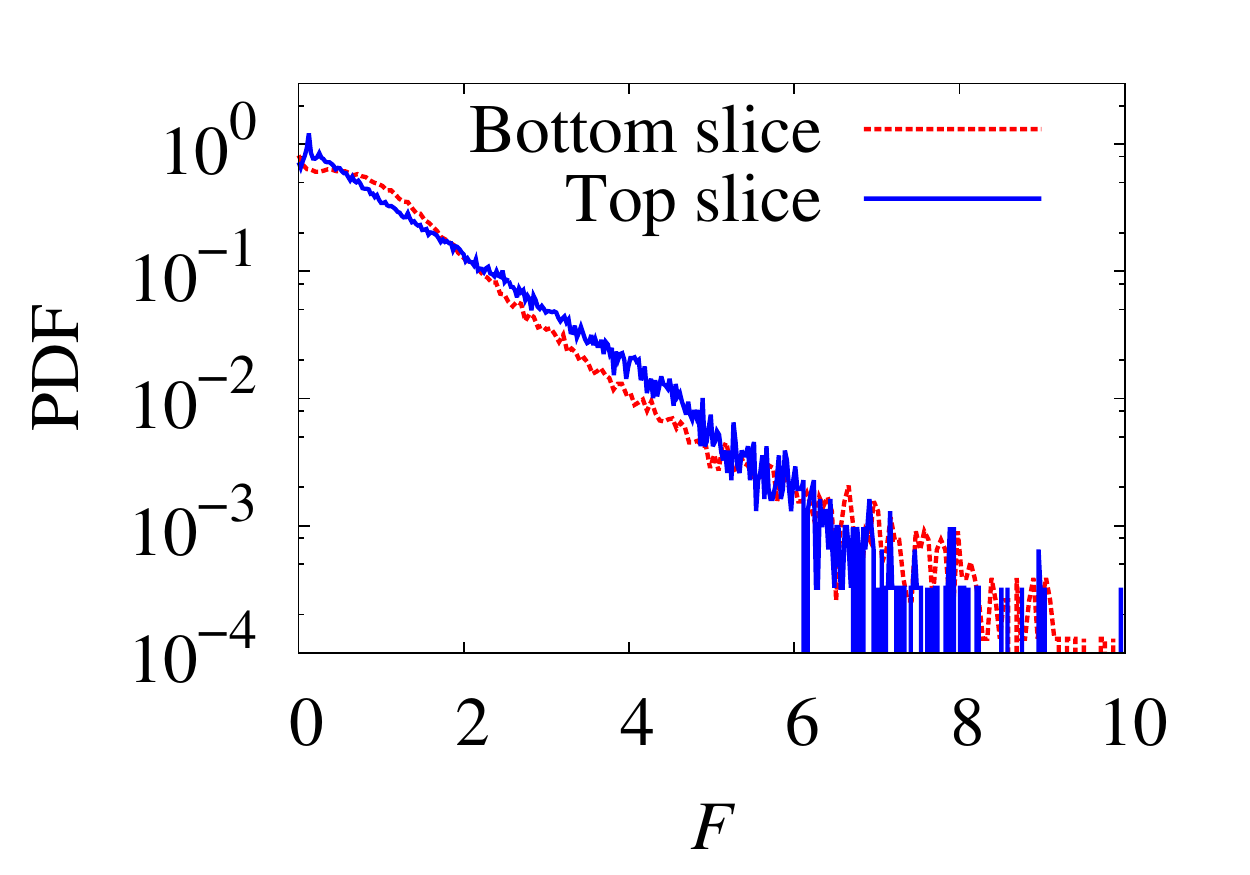}}
\subfigure[Tangential forces.]
{\includegraphics[width=0.49\columnwidth]{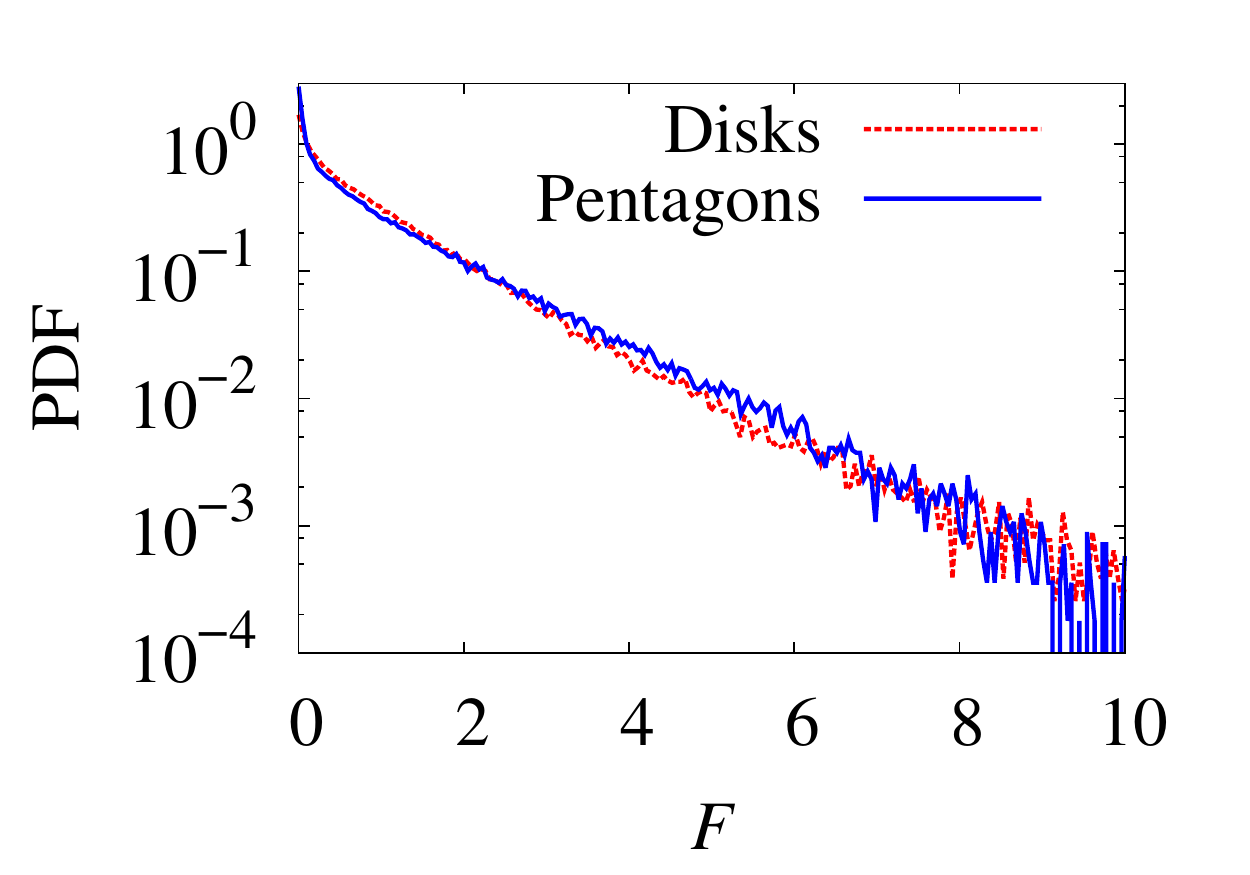}}
\caption{(Color online) Probability density functions (PDF) (disks, low tapping). }
\label{fig:disks_high_pdf}
\end{figure}

Figure~\ref{fig:disks_high_pdf} shows PDFs of normal and tangential forces for the top and the bottom slice of the disk-based system. The comparison of the PDFs for different slices does not reveal any prominent differences on a statistical level. The only noticeable difference can be found at low normal forces where the PDF of the bottom slice shows a plateau. This behavior is expected since gravity leads to compression, and it is known that the PDFs for granular  systems exposed to stronger compression show a plateau at small forces, see e.g.,~\cite{pre14}.

\begin{figure}
\centering
\subfigure[Normal forces.]{\includegraphics[width=1.6in]{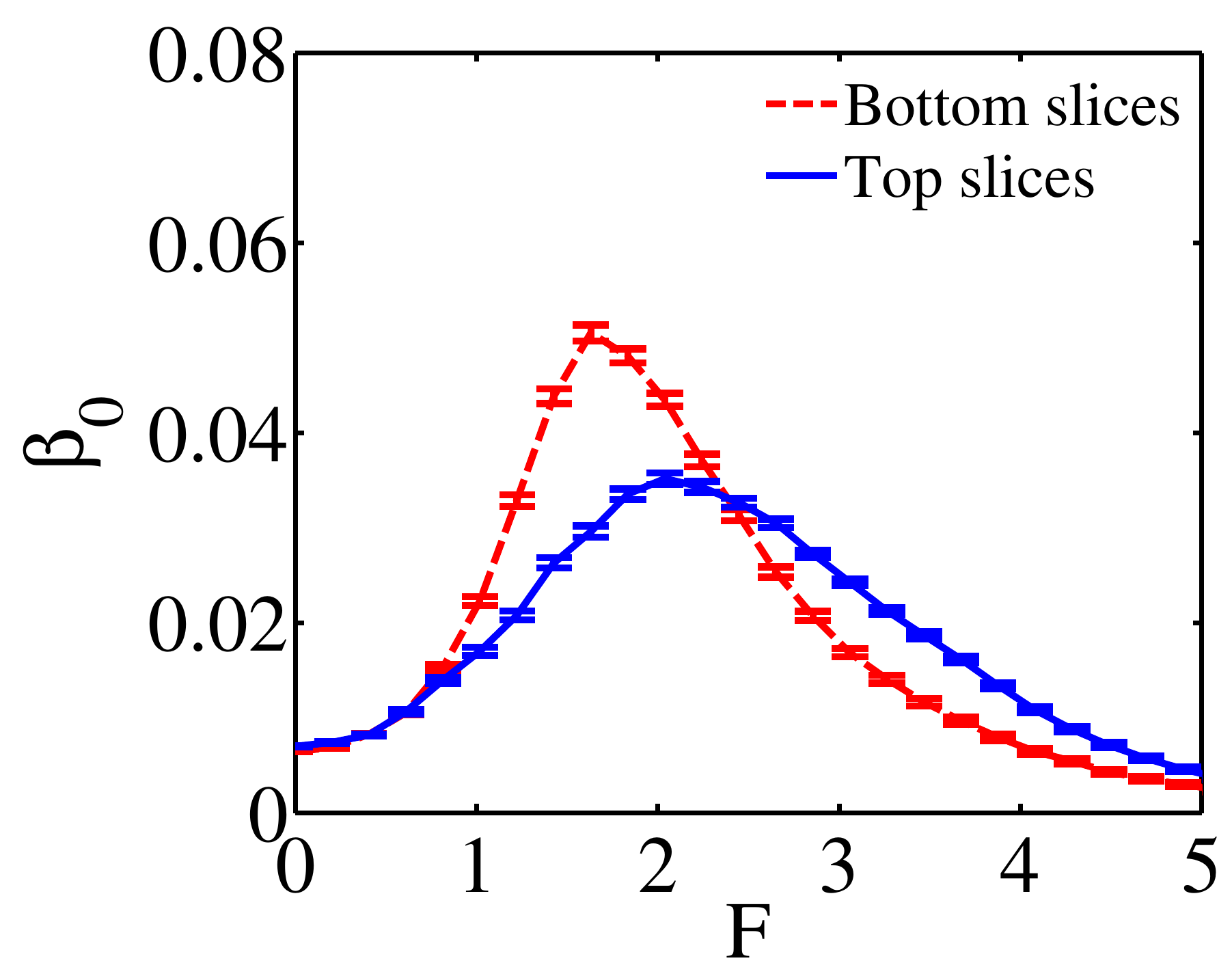}}
\subfigure[Tangential forces.]{\includegraphics[width=1.6in]{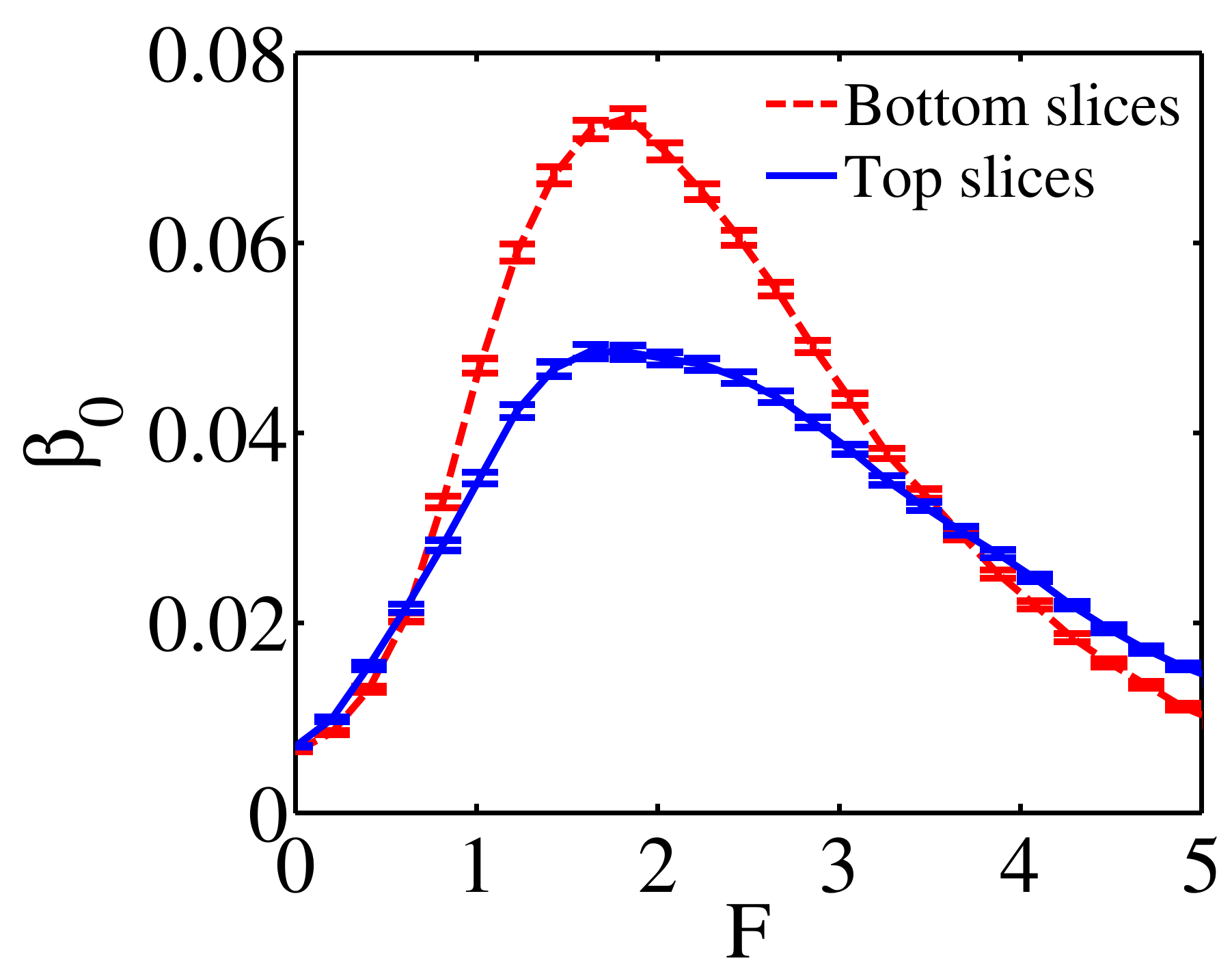}}
\caption{(Color online) $\beta_0$ (disks, low tapping).   Error bars correspond to the standard error estimated as 
$std/\sqrt{N}$ with $std$ the standard deviation and $N=500$, the number of realizations considered.}
\label{fig:disks_top_bottom}
\end{figure}

Now we turn our attention to structural properties of the force networks for the two slices. We investigate changes in the number of  components, present in the force network, as we change the force threshold. Figure~\ref{fig:disks_top_bottom} shows $\beta_0$, as a function of force threshold for the top and bottom slices. For small force threshold, $F$, all particles are connected and $\beta_0 =1$. If $F$ is large, then there are very few contacts experiencing a force larger than $F$. This leads to a small number of clusters (small value of $\beta_0$). As shown in Fig.~\ref{fig:disks_top_bottom}, the number of  components reaches the maximum for an intermediate value of $F$. This value is similar for the top and bottom slices ($F\approx 2$). However, the maxima differ significantly. For normal forces there is a much larger number of components in the bottom slice for $F \in [1,2]$. For  the tangential forces this is true even for a wider range of $F$.

Due to  gravitational compaction, the average pressure is larger in the bottom slice. We conclude that this increase in the pressure leads to the formation of a larger number of isolated components and increases the ramification of the force network. This finding is consistent with the results obtained for isotropically compressed systems, where the peak of $\beta_0$ curves was shown to increase with compression~\cite{epl12}.  

While additional information could be extracted from other measures, the number of components as a function of force threshold already shows that the properties
of force networks are depth-dependent. Therefore, in the rest of this paper, we focus only on the bottom slices, so that various comparisons 
are not obscured by heterogeneity.

\subsection{Structural differences in force networks for the systems exposed to
different tap intensities}
\label{sec:high_low}

In this section, we attempt to distinguish properties of the force networks corresponding to different $\Gamma$s that 
lead to the same steady state packing fractions for disks. The mean stress tensor differs for the steady states at 
different $\Gamma$~\cite{pugnaloni_pre10}. Therefore, the states are distinguishable through macroscopic properties.  
However, our goal is to distinguish the  force networks after normalizing by  the average force.

\begin{figure}[]
\centering
\subfigure[Normal forces.]
{\includegraphics[width=0.49\columnwidth]{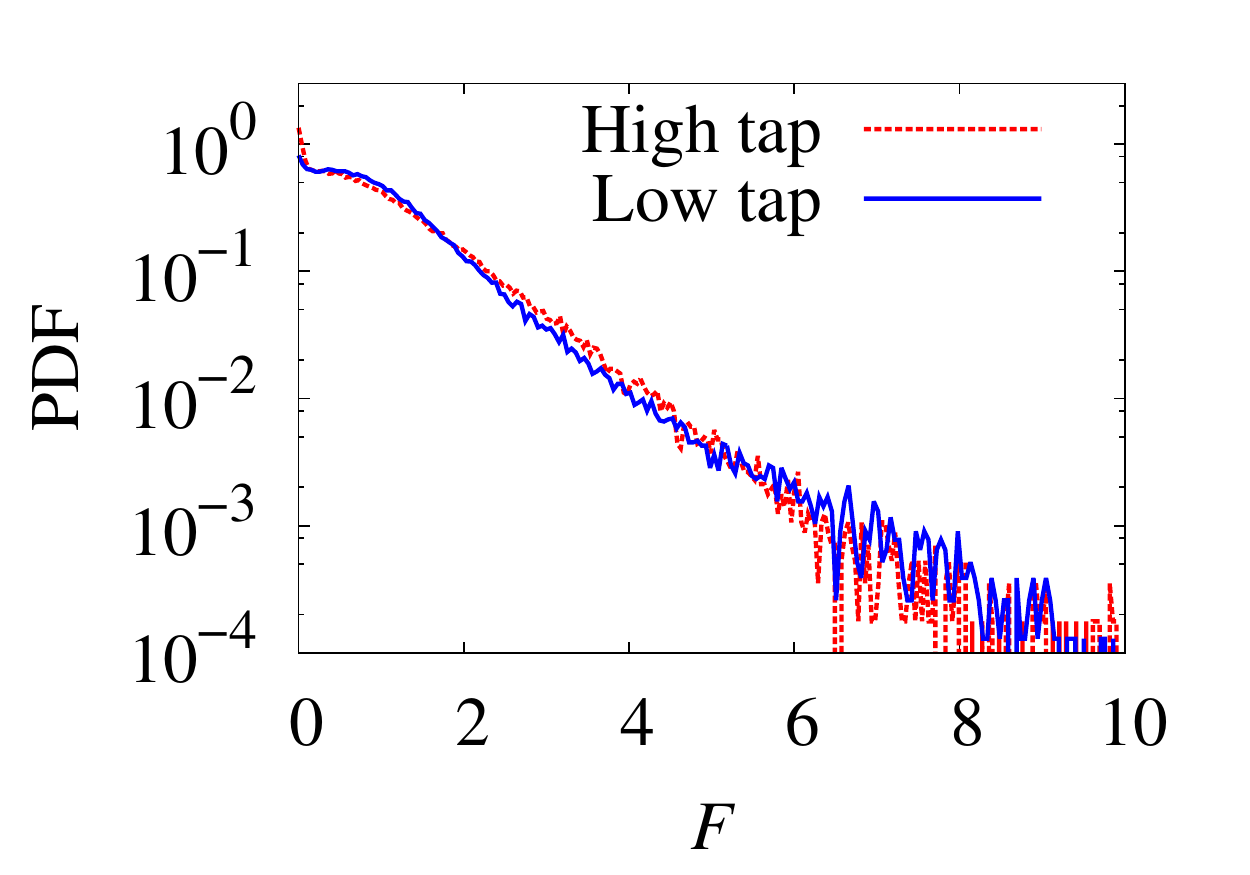}}
\subfigure[Tangential forces.]
{\includegraphics[width=0.49\columnwidth]{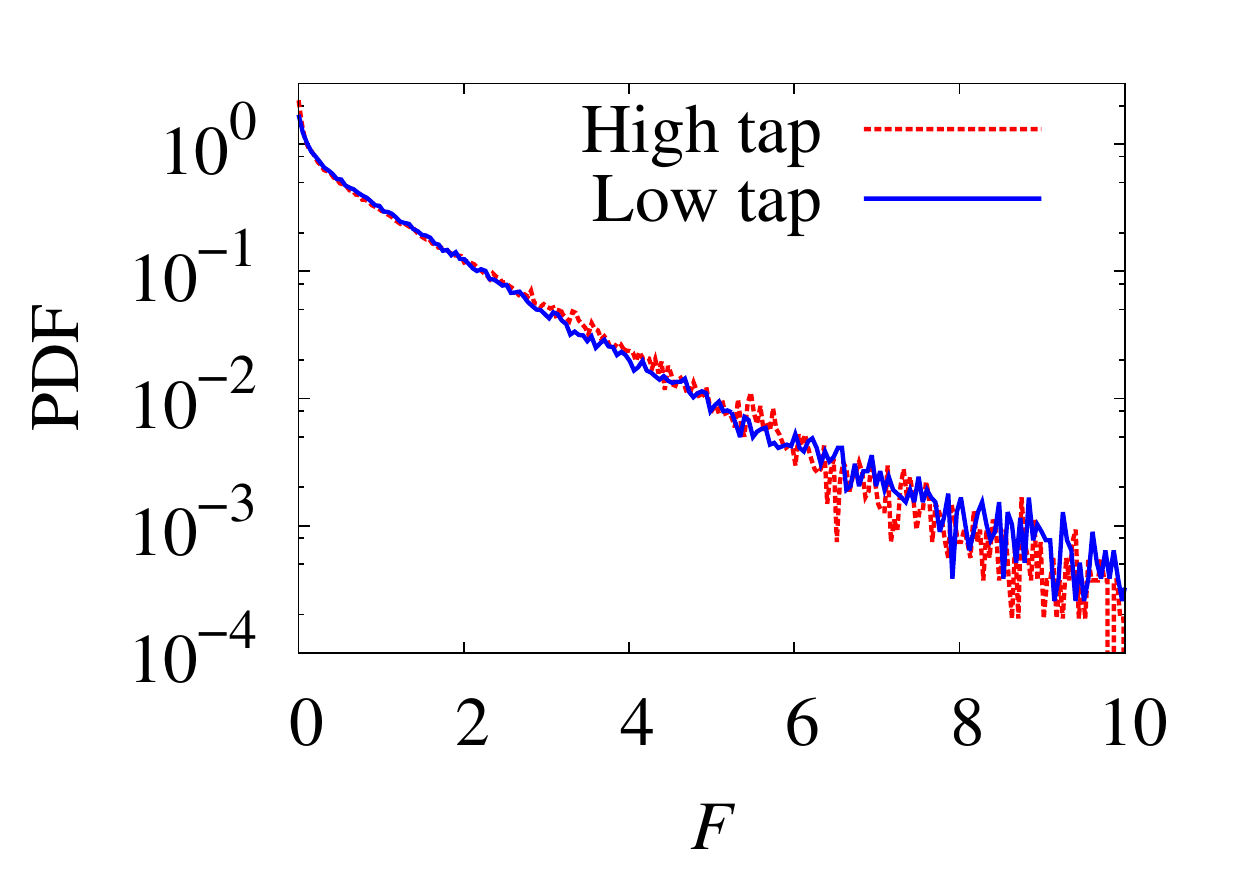}}
\caption{(Color online) PDFs (disks, bottom slice).}
\label{fig:disks_bottom_pdf}
\end{figure}

Figure~\ref{fig:disks_bottom_pdf} shows the PDFs for the systems composed  of disks, for both normal and tangential forces. Note that they do not show any significant differences. This is consistent with  the results of similar simulations reported in~\cite{carlevaro2012arches}. The PDFs are unaffected by the value of $\Gamma$ at least as long as they lead to the same $\phi$.  

We note that for low tap intensity, the particles barely move, while for 
high tapping intensity the particles completely rearrange between the consecutive taps. Recalling 
further that we are considering monodisperse disks, one could expect that a certain degree of 
ordering/crystallization occurs in the system tapped at low intensity. This may lead to long-range correlation in the contact forces that may complicate the comparison with disordered structures such as the ones observed in polydisperse systems or in packings of non-circular grains.    

\begin{figure}
\includegraphics[width=0.9\columnwidth]{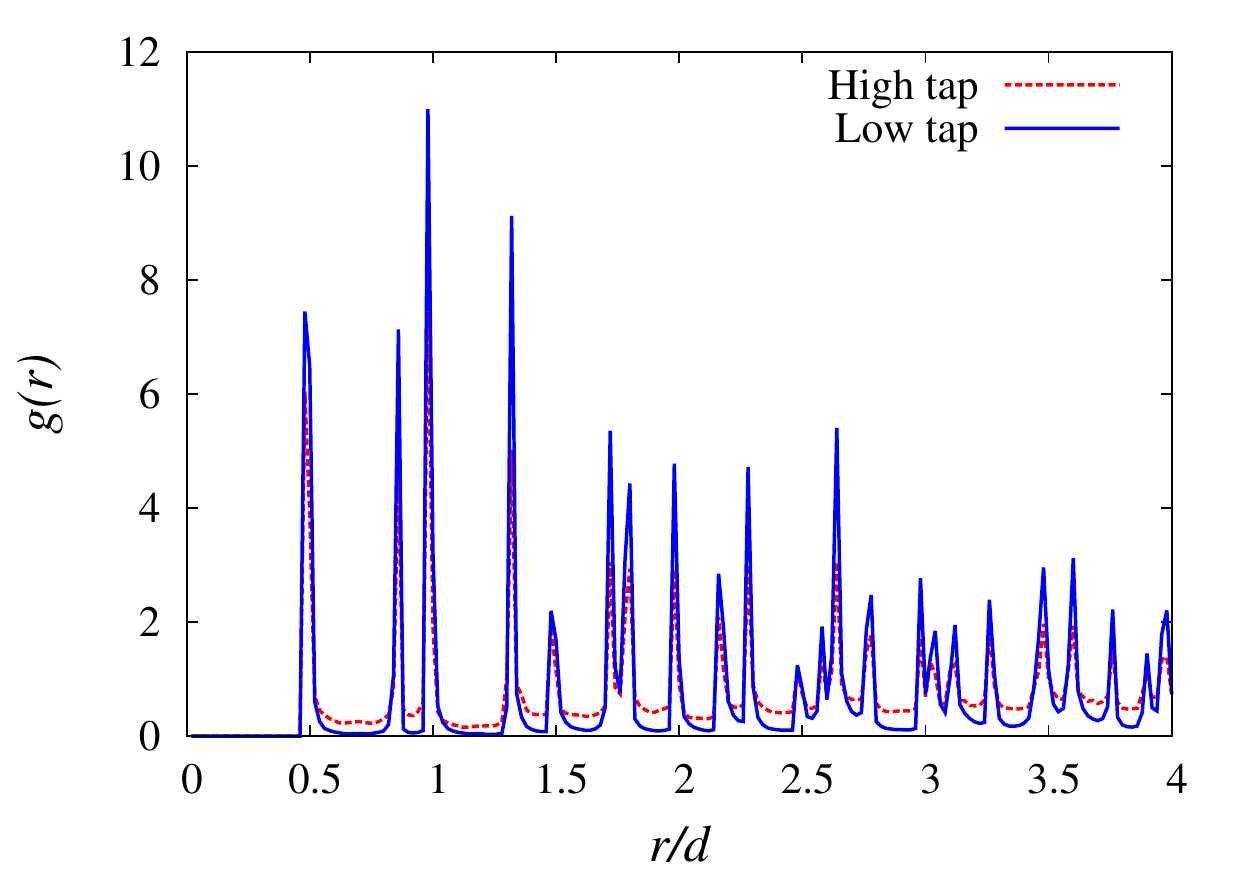}
\caption{(Color online) Pair correlation function, $g(r)$ (disks, bottom slice).}\label{fig:gofr}
\end{figure}

Figure~\ref{fig:gofr} shows the contact--contact pair correlation function, $g(r)$. The $g(r)$ is calculated as usual, see e.g.~\cite{allen87}, but using the position of contact points instead of the centers of the particles. To avoid boundary effects on $g(r)$ due to the empty spaces outside the containing box, we only average over particles in a central area of the slice, away from the walls. Figure~\ref{fig:gofr} shows that there are indeed long-range correlations in the disk packings, and these correlations are, as expected, much stronger for low intensity taps. The first three peaks in the $g(r)$ correspond to the three typical distances between the six contacts of a circular grain in a triangular lattice arrangement.     

To asses the impact of spatial ordering on the properties of force distribution, we consider the correlation of the contact forces by calculating the force--force correlation function, $f(r)$, defined as  
\[ 
f(r) = \frac{1}{\rho_c}\sum\sum_{i,j>i}{\delta(r-r_{i,j})f_if_j},
\]  
where $\rho_c$ is the density of contact points, $\delta$ is the Dirac delta distribution, $r_{i,j}$ is the contact-to-contact distance and $f_i$ is the force experienced at the contact $i$, normalized by the average force $\langle f \rangle$. The sum runs over all pairs of contacts.

\begin{figure}
\includegraphics[width=0.9\columnwidth]{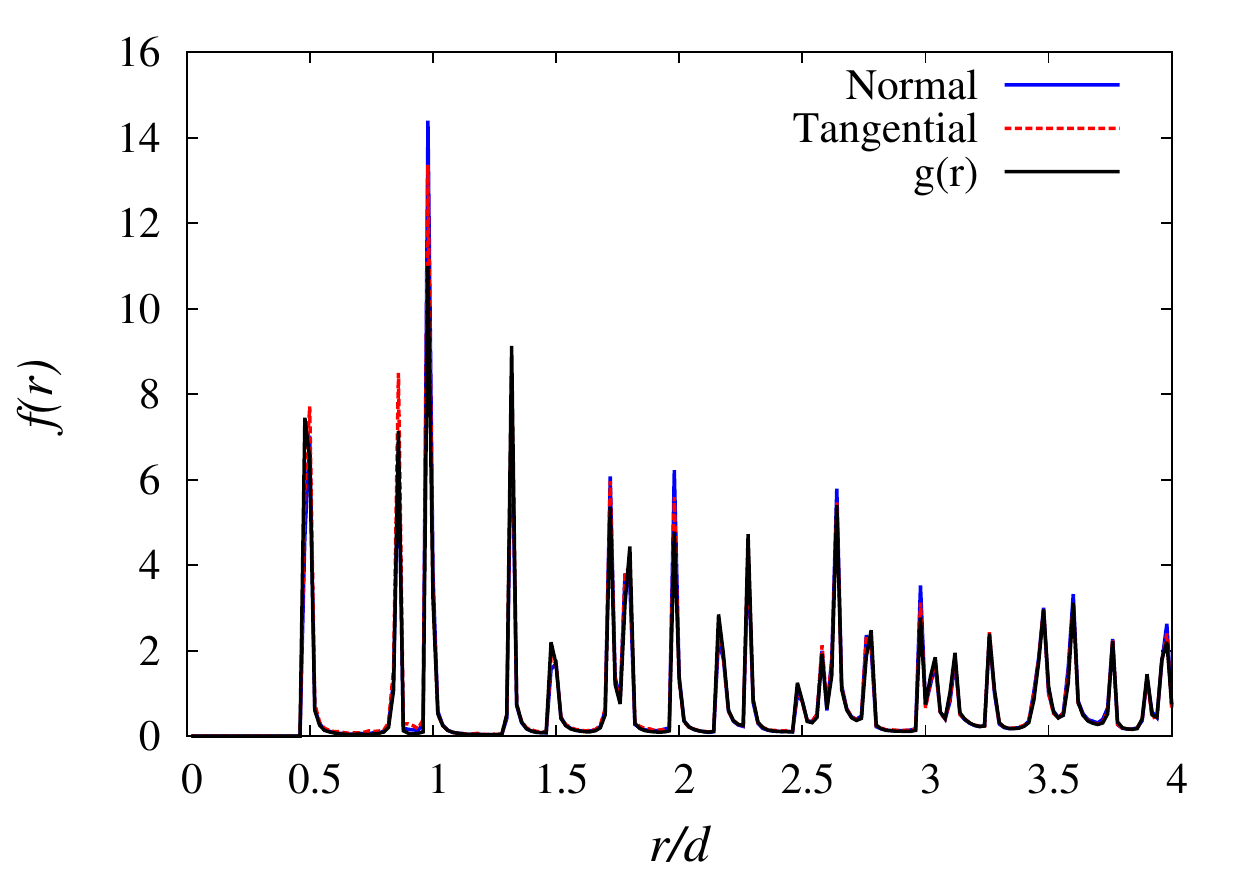}
\caption{(Color online) Force correlation function, $f(r)$ (disks, low tapping).  The $g(r)$ is also shown. \label{fig:f} }
\end{figure} 

\begin{figure}
\centering 
\subfigure[Normal forces.]
{\includegraphics[width=0.49\columnwidth]{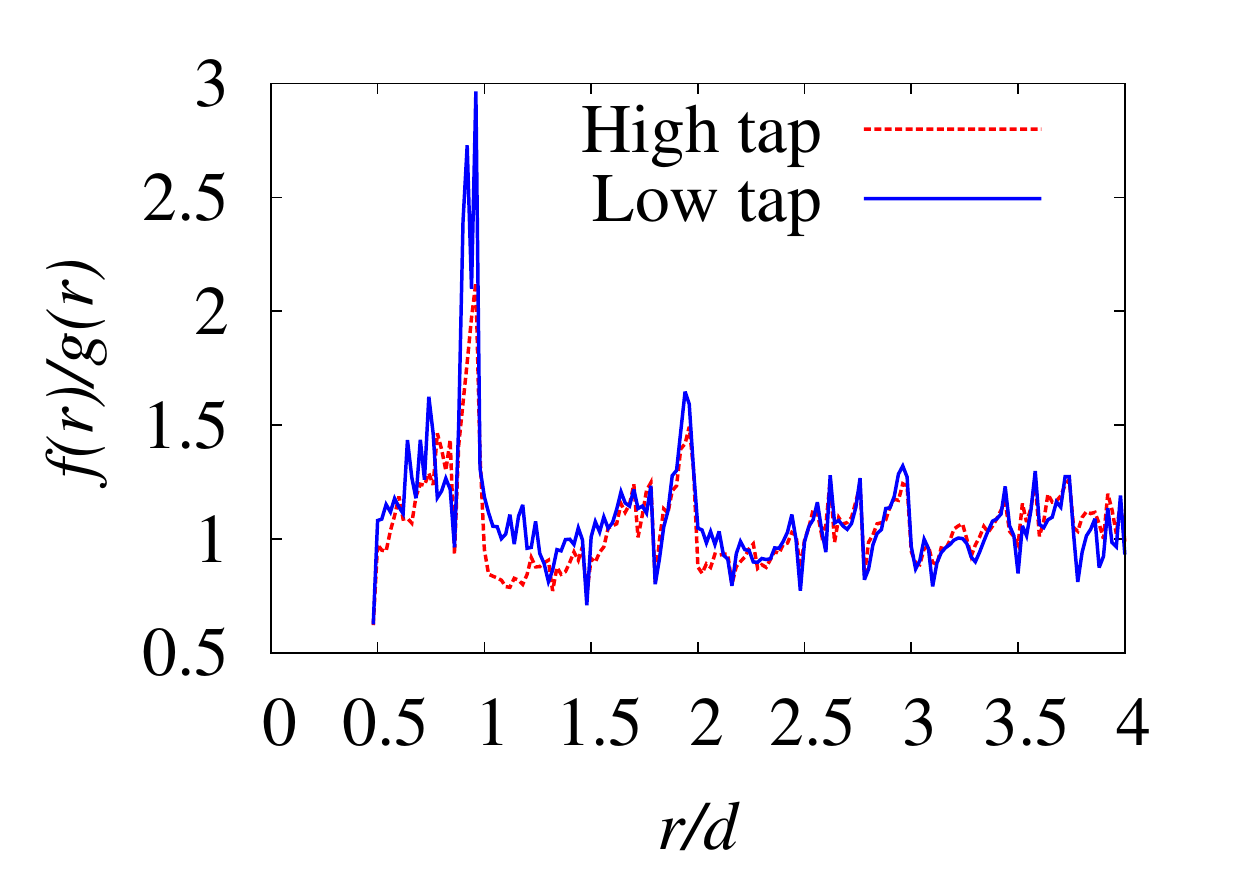}}
\subfigure[Tangential forces.]
{\includegraphics[width=0.49\columnwidth]{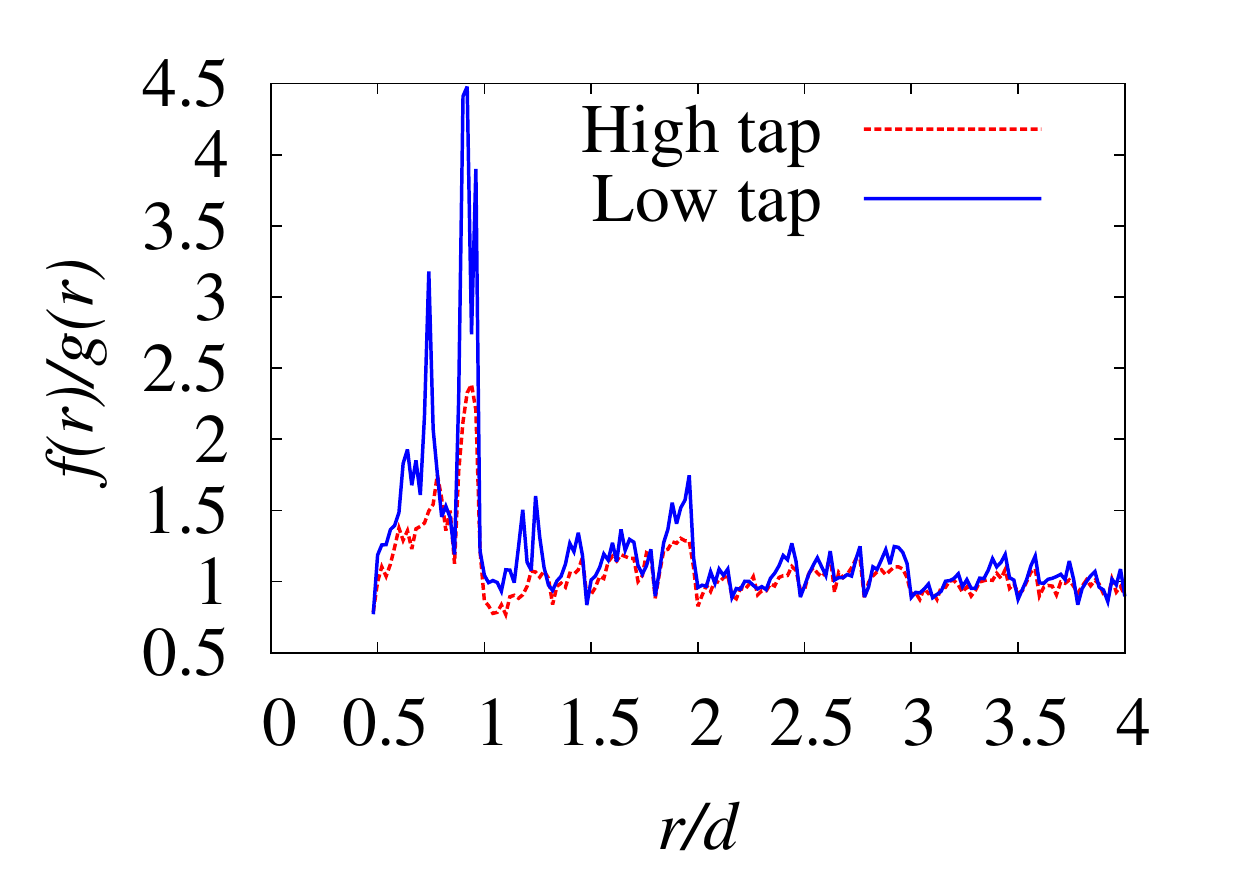}}
\caption{(Color online) $f(r)/g(r)$ (disks, bottom slice).}\label{fig:normalized}
\end{figure}

Figure~\ref{fig:f} shows $f(r)$ for disks for the low tap intensity.  The overall shape of $f(r)$ agrees with $g(r)$ since the main correlation comes from the positional order of the contact points. To highlight the contribution due to the strength of the forces, we consider the ratio $f(r)/g(r)$, shown in Fig.~\ref{fig:normalized},  for both low and high tap intensities. There is much less structure in the force correlation if the positional correlations are eliminated. Besides, there is no substantial difference between the considered low and high tap intensities. Therefore, positional ordering (crystallization) does not reflect itself in an ordering of the force strengths. The only strong peak in $f(r)/g(r)$, at $r/d=1$, indicates that two contacts on a single grain that are acting from opposite sides have a large probability of having strong forces.  However, this is only a short-range correlation.   
 
\begin{figure}
\centering
\subfigure[Normal forces.]{\includegraphics[width=1.6in]{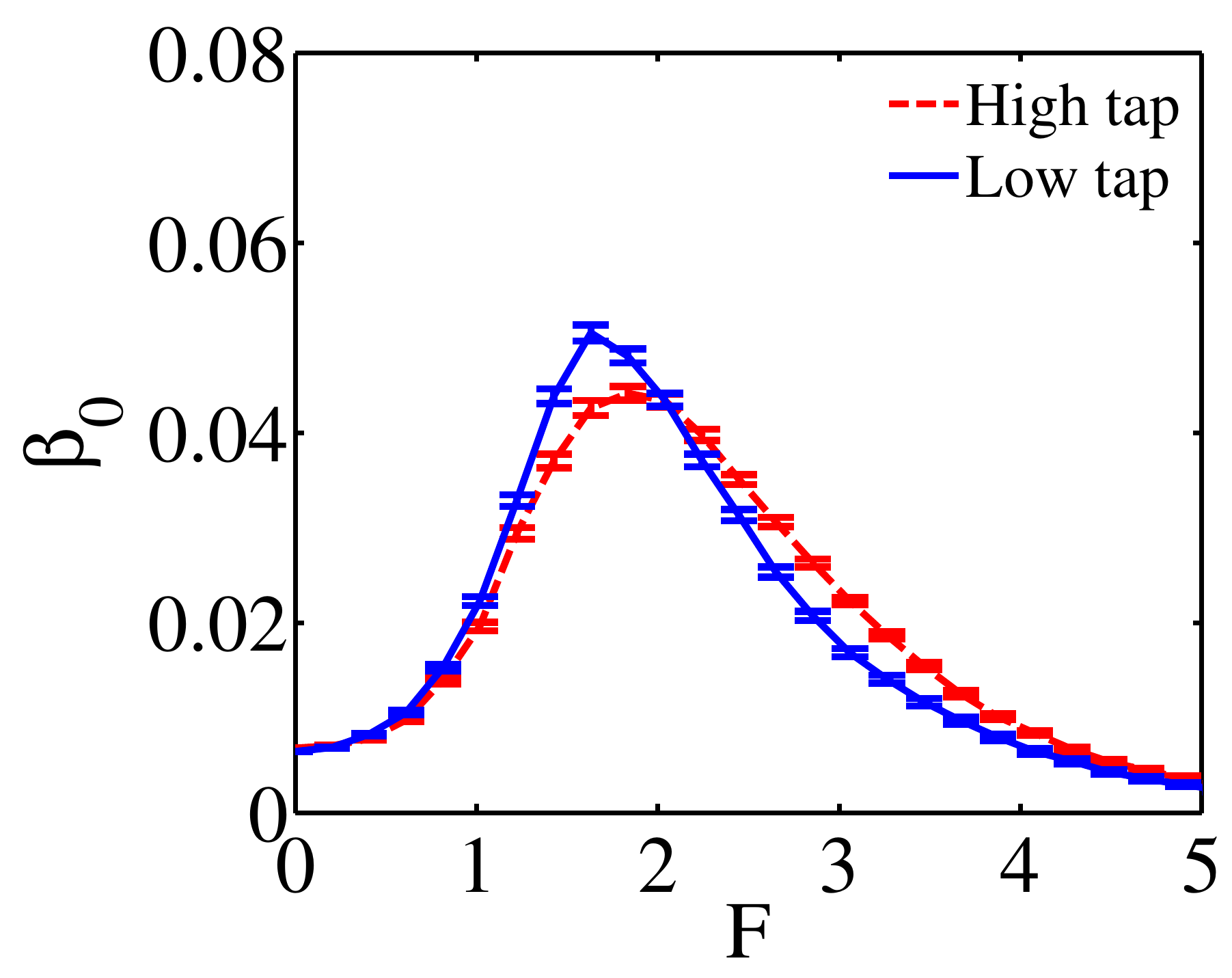}}
\subfigure[Tangential forces.]{\includegraphics[width=1.6in]{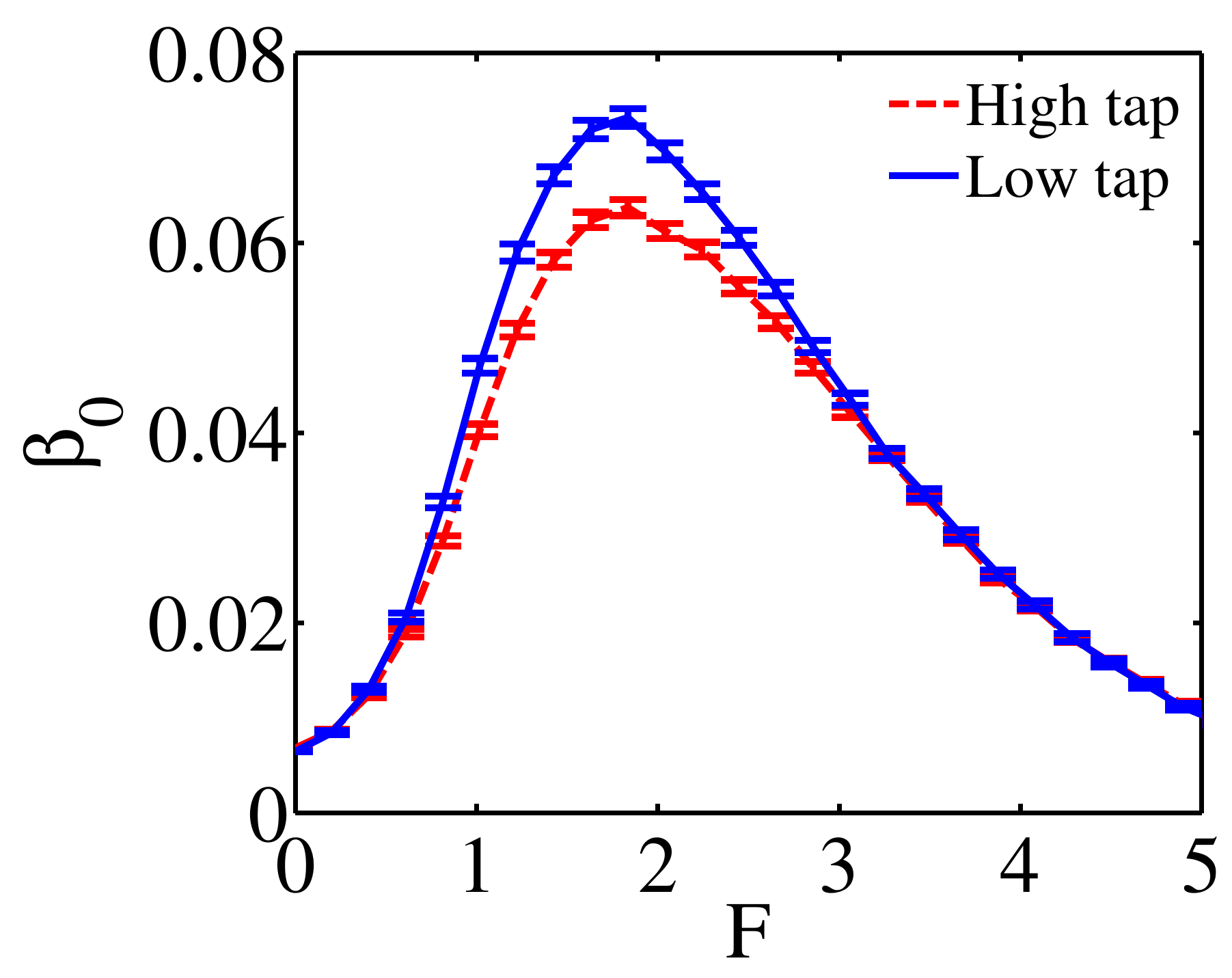}}
\subfigure[Normal forces.]{\includegraphics[width=1.6in]{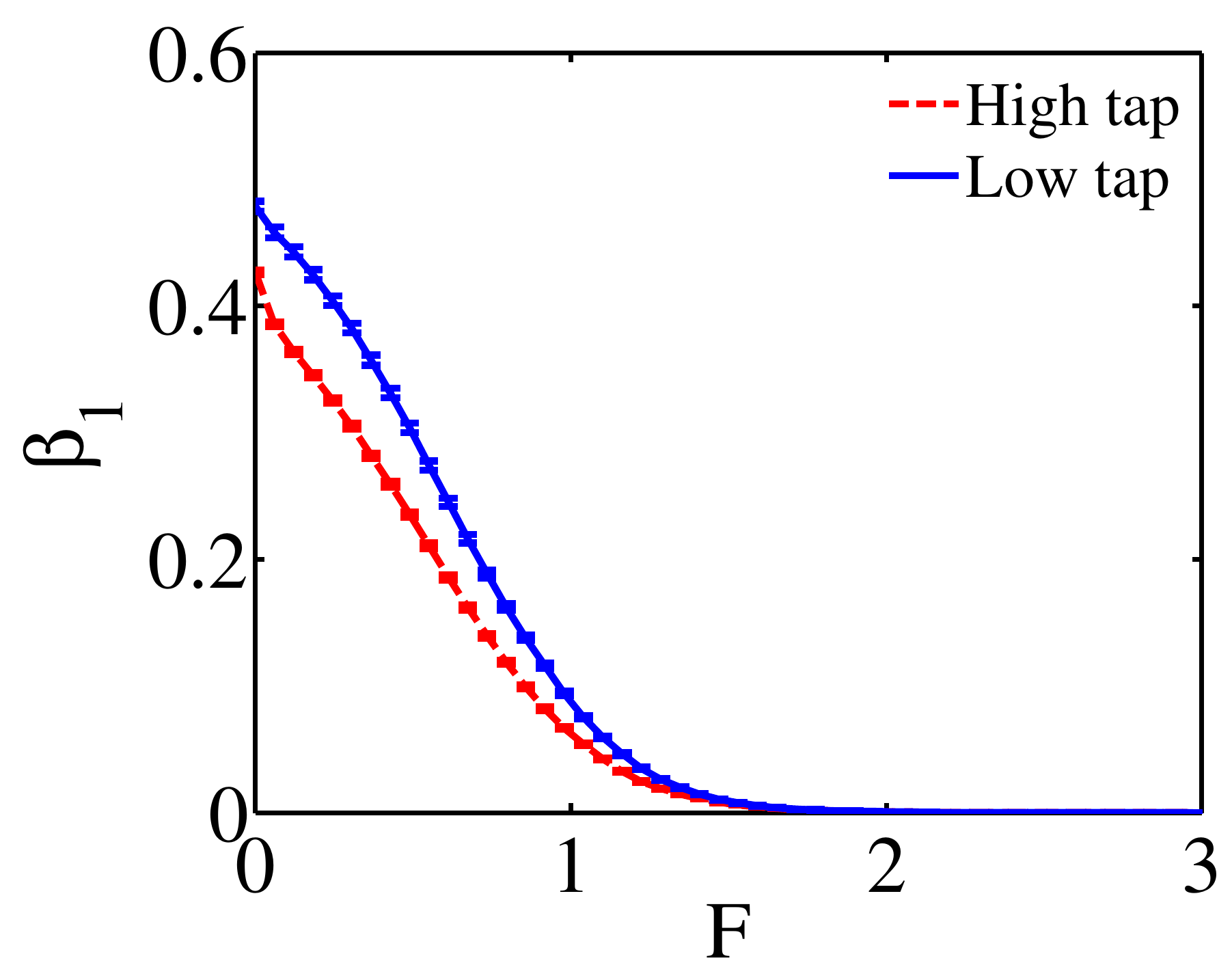}.pdf}
\subfigure[Tangential forces.]{\includegraphics[width=1.6in]{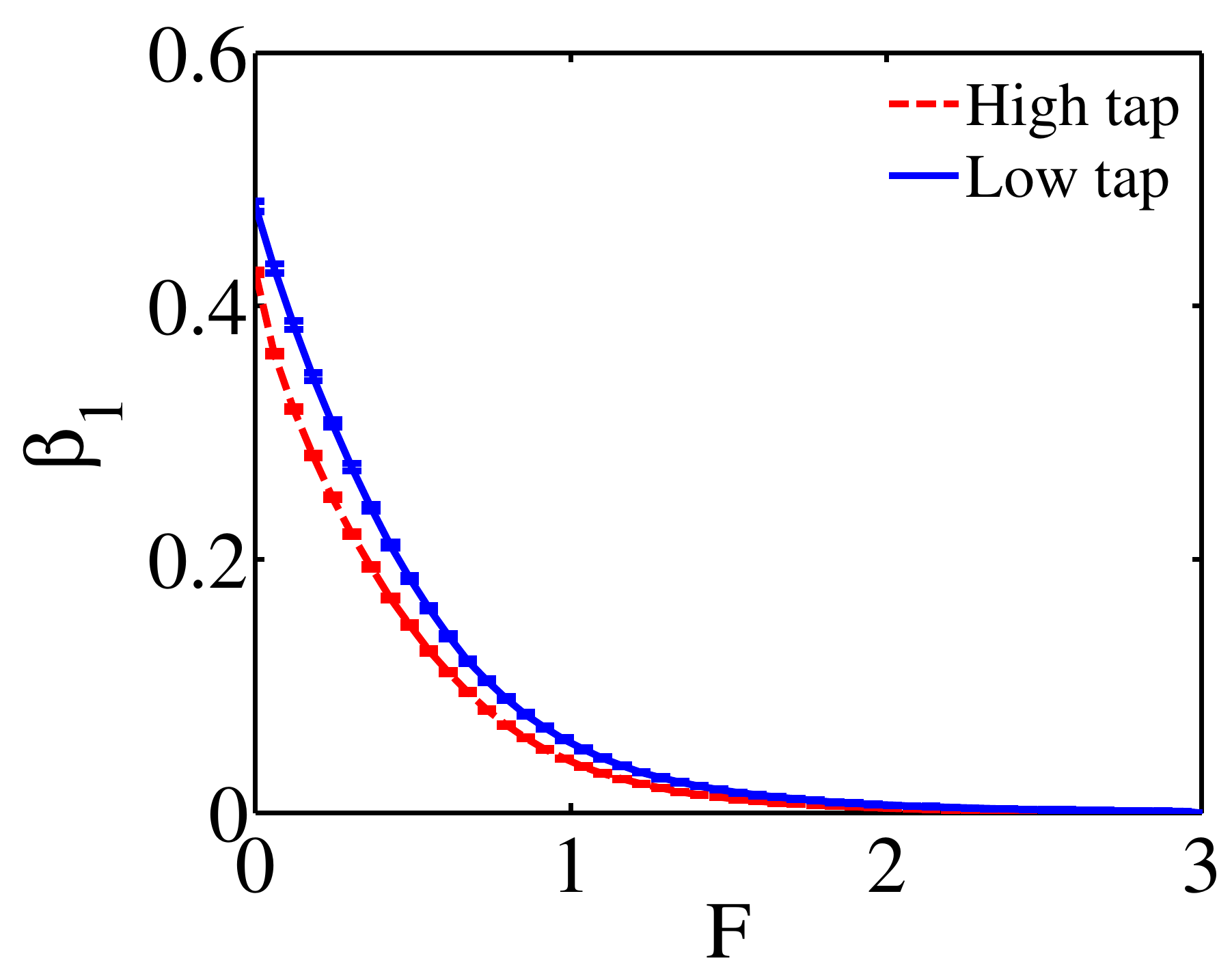}}
\caption{(Color online)  (a) and (b) $\beta_0$, and (c) and (d) $\beta_1$  (disks, bottom slice). Error bars are as in Fig. \ref{fig:disks_top_bottom}.}
\label{fig:disks_low_high}
\end{figure}

Now we compare structural differences of the force networks corresponding to the systems prepared with different tap intensities. As can be seen in Fig.~\ref{fig:disks_low_high},  counting the number of  components, $\beta_0$, and the number of loops, $\beta_1$, present in different networks, does not reveal any remarkable difference between these regimes. For brevity, we show only the count of non-trivial loops. However, including the
trivial, three-particle loops does not distinguish the systems either (figures not shown for brevity).

In summary, strong correlations of particle positions do not necessarily lead to correlations
of the forces between the particles. Furthermore, none of the measures considered (PDFs, force correlation
functions, $\beta_0$, $\beta_1$) uncover any significant differences between the networks of steady states characterized
by the same  $\phi$ but obtained using different tap intensities.  It is worth mentioning that previous simulations and experiments \cite{arevalo_pre13,ardanza2014topological} have found some differences in the contact networks (i.e., the force network with $F=0$) by considering the different loop sizes (trivial loops seem to be numerous in the high tap states). We notice that in the current work walls present virtually zero friction whereas previous studies were done with frictional walls. Also, we focus on the properties of a horizontal layer of grains in contrast with the analysis of the entire column done previously. Based on our results, we conclude that a more careful analysis of the structure of the force networks is needed for the purpose of finding the differences between these equal-$\phi$ states. Such analysis is presented in~\cite{paper2}. 

\subsection{Structural differences of the force networks for disks and pentagons}
\label{sec:disks_pents}

Understanding differences between force networks in the systems composed of particles of different shapes is important to asses to what extent simple models based on circular/spherical grains capture the properties of realistic systems. 

Earlier studies of the contact forces for elongated particles \cite{hidalgo2009role,hidalgo2010granular}, pentagons \cite{azema2007force} and irregular polyhedral particles \cite{radjai_09} revealed  that the orientational distribution of the contact forces is more anisotropic than for circular or spherical grains. However, this descriptor has a local character and little is known about the general properties of the underlying force networks. In this section, we compare different systems using the force PDFs and structural properties of the force networks. These methods provide complementary insights  regarding the differences between the considered systems.  For brevity, we consider only  the low tapping regime.

\begin{figure}[]
\centering
\subfigure[Normal forces.]{\includegraphics[width=0.49\columnwidth]{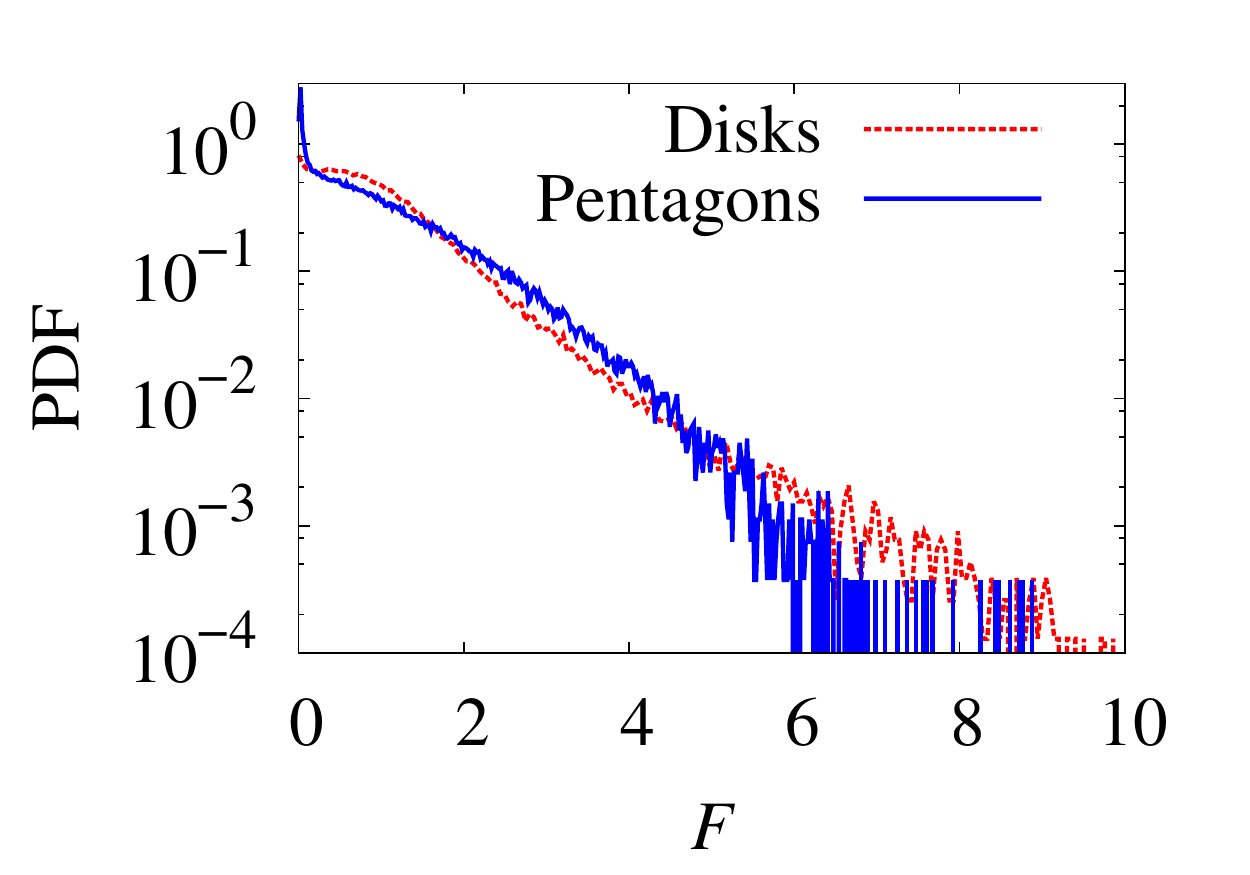}}
\subfigure[Tangential forces.]{\includegraphics[width=0.49\columnwidth]{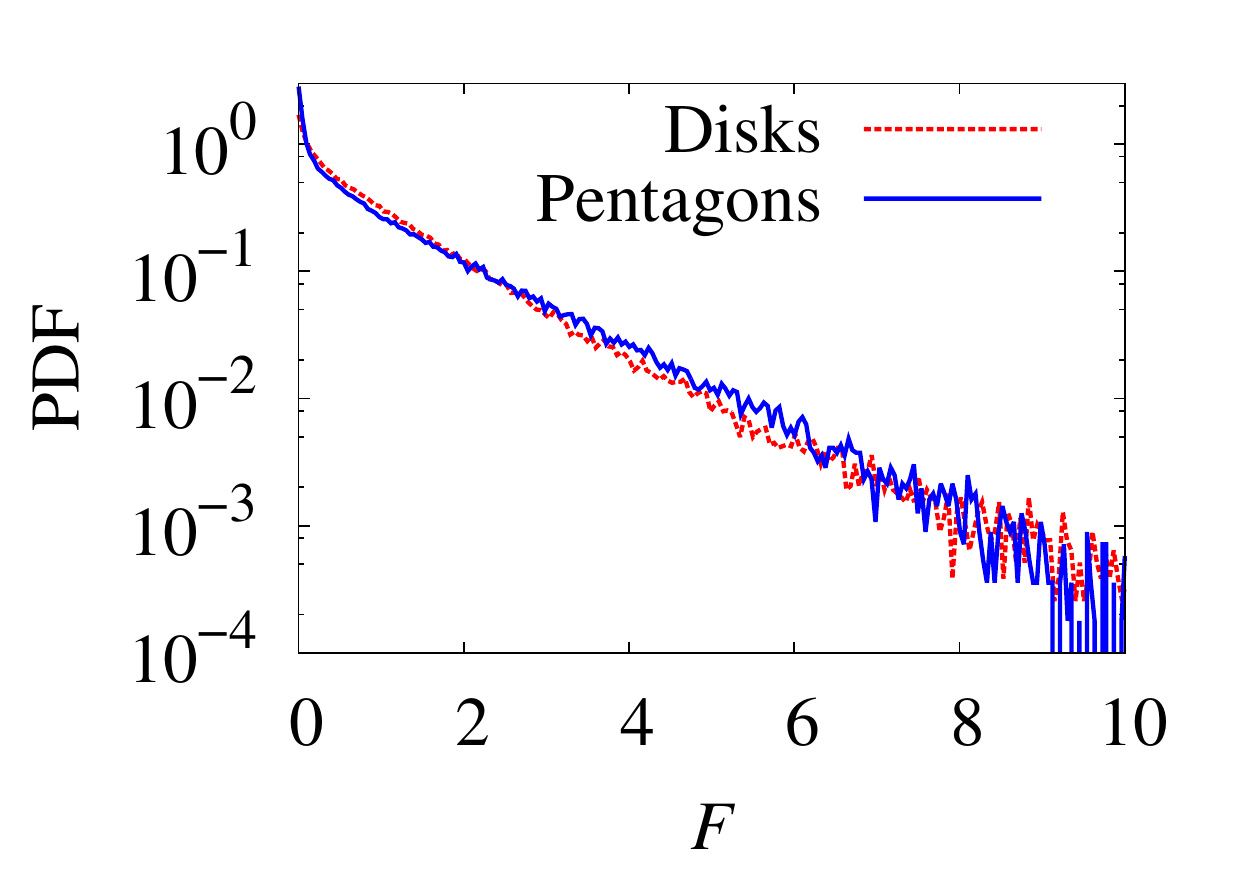}}
\caption{(Color online) PDFs (bottom slice, low tapping).}
\label{fig:disks_pents}
\end{figure}

We start by considering the PDFs for disks and pentagons, shown in Fig.~\ref{fig:disks_pents}. Note that the distributions of the tangential forces are indistinguishable. On the other hand, there is  a faster decay at large normal forces for pentagons.  Disks contain a larger number of contacts than pentagons at normal forces around $F = 1$. This is in  agreement with the results for similar systems exposed to isotropic compression \cite{azema2007force}. Let us now consider the behavior of the PDF of the normal forces  for $F > 1$. For pentagons, the decay  is consistent with Gaussian behavior while for the disks the decay is exponential.   A Gaussian decay has been shown to be connected with the presence of arches in the structure \cite{carlevaro2012arches}. This is consistent with the intuition that pentagons are more prone to form arches than circular grains. However, the differences in PDF are rather subtle. Moreover, disk packings can also show Gaussian-like PDF depending on the preparation protocol \cite{makse_prl99}; hence a Gaussian PDF is not a signature of non-circular particles. 
In the rest of this section, we show that more significant differences between the systems can be identified by considering the global geometry of the force networks.  

\begin{figure}
\centering
\subfigure[Normal forces.]{\includegraphics[width = 1.6in]{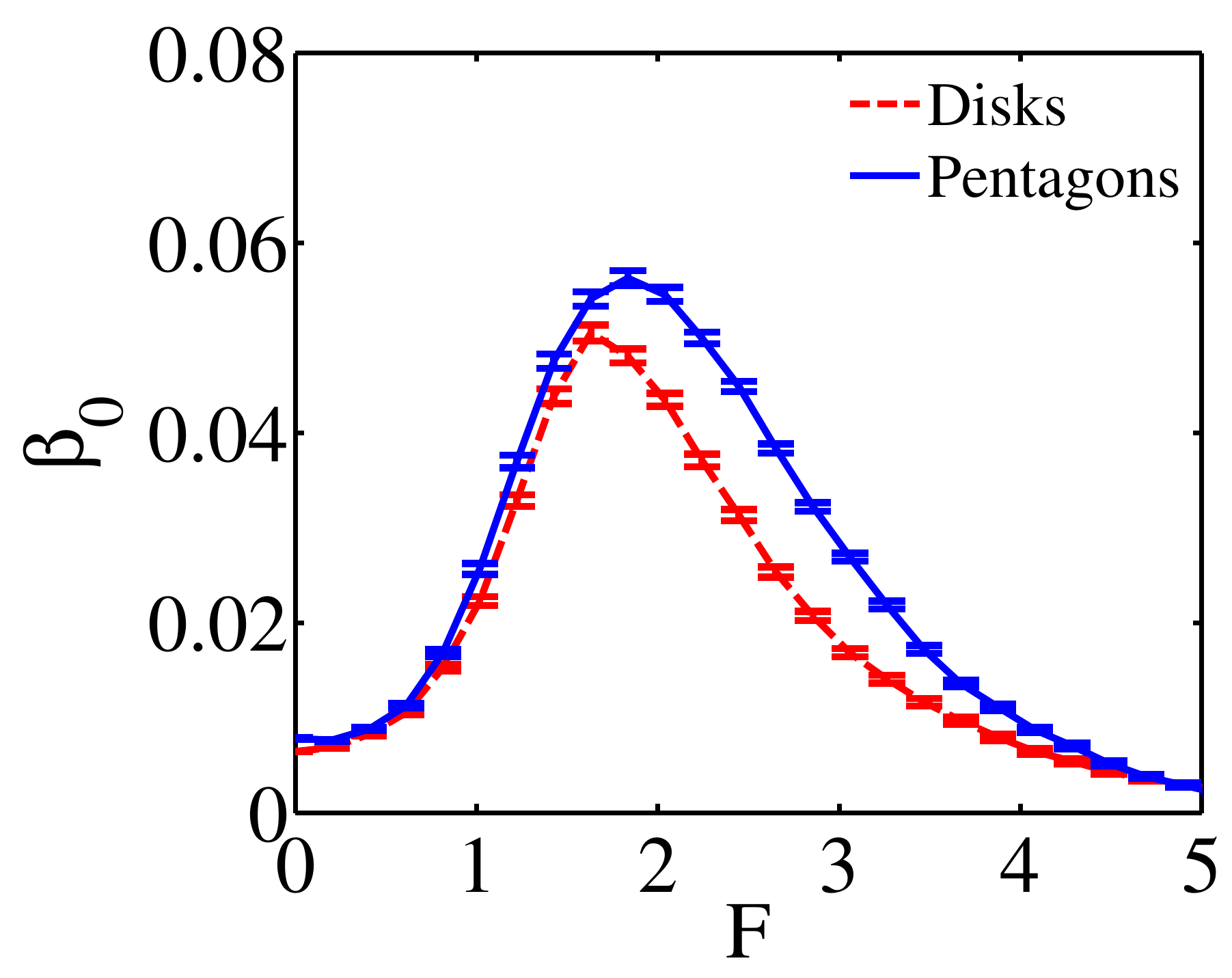}}
\subfigure[Tangential forces.]{\includegraphics[width = 1.6in]{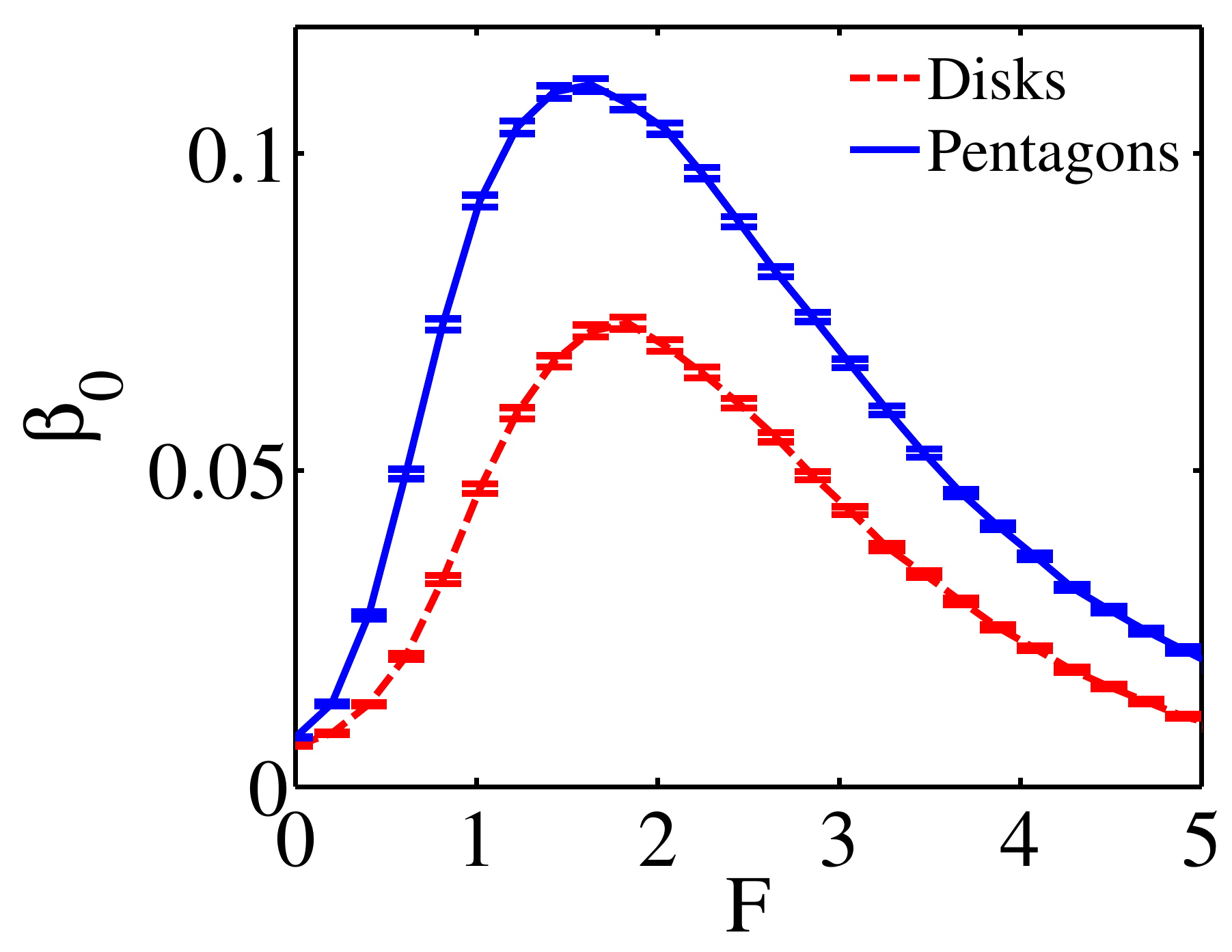}}
\caption{(Color online) $\beta_0$ (bottom slice, low tapping).}
\label{fig:disks_pents_b0}
\end{figure}

Figure~\ref{fig:disks_pents_b0} shows $\beta_0$ as a function of the force threshold, $F$, for disks and pentagons. 
For normal forces, $\beta_0$  is rather similar for both types of particles.  In the range $2 < F < 4$, the number of components is slightly larger for pentagons.   
This shows a correlation with the PDFs given in Fig.~\ref{fig:disks_pents}(a) where pentagons show a larger number of contacts in the same range. 
Note that these features do not need to be correlated since $\beta_0$ depends not only on the number of contacts at a particular force level, 
but also on how these contacts (edges) connect with the edges in the portion of the force network that corresponds to larger $F$. 
In general, $\beta_0$ is larger for tangential  than for normal forces [compare Figs.~\ref{fig:disks_pents_b0}(a) and (b)].   
This indicates that the edges of the normal force network corresponding to the contacts characterized by strong  forces tend to be more connected and form less clusters than for the tangential force network. Hence, for the normal force network, the edges with lower force values are more likely to be connected to the components present in the force network for larger force threshold. 
This effect is observed in Fig.~\ref{fig:snap}, where force chains (corresponding in the loose sense to the  components at high $F$) 
are larger for the normal than for the tangential forces.

Restricting our attention to the tangential forces, Fig.~\ref{fig:disks_pents_b0}(b) indicates that $\beta_0$ is significantly larger for the pentagons than for disks.
This means that, as we reduce the threshold $F$ from a large initial value, the new contacts that come into play in the tangential force network tend to be more disconnected from the previous high force contacts for pentagons, in comparison to disks.

\begin{figure}
\centering
\subfigure[$F=3.0$]{\includegraphics[width = 1.6in]{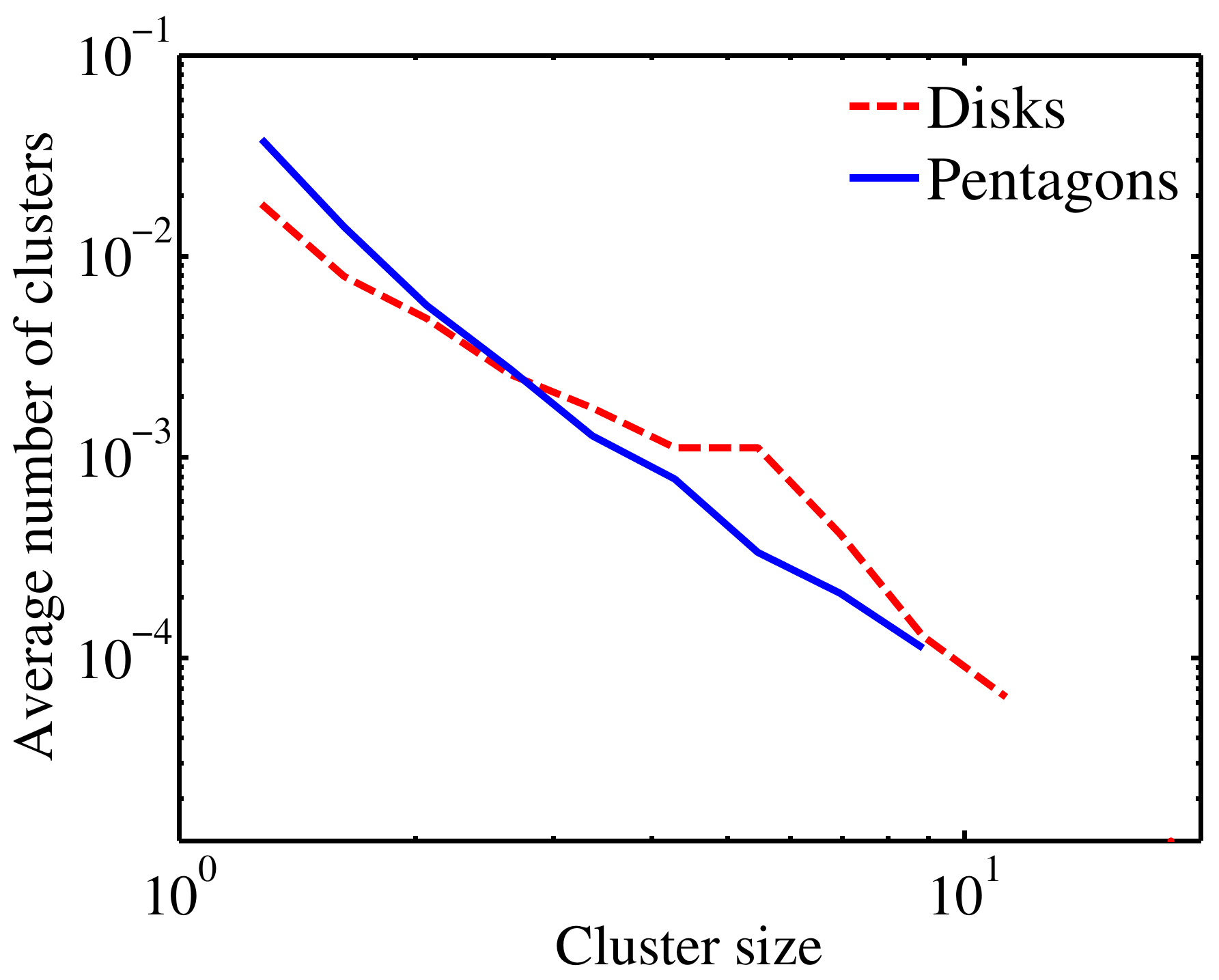}}
\subfigure[$F=2.0$]{\includegraphics[width = 1.6in]{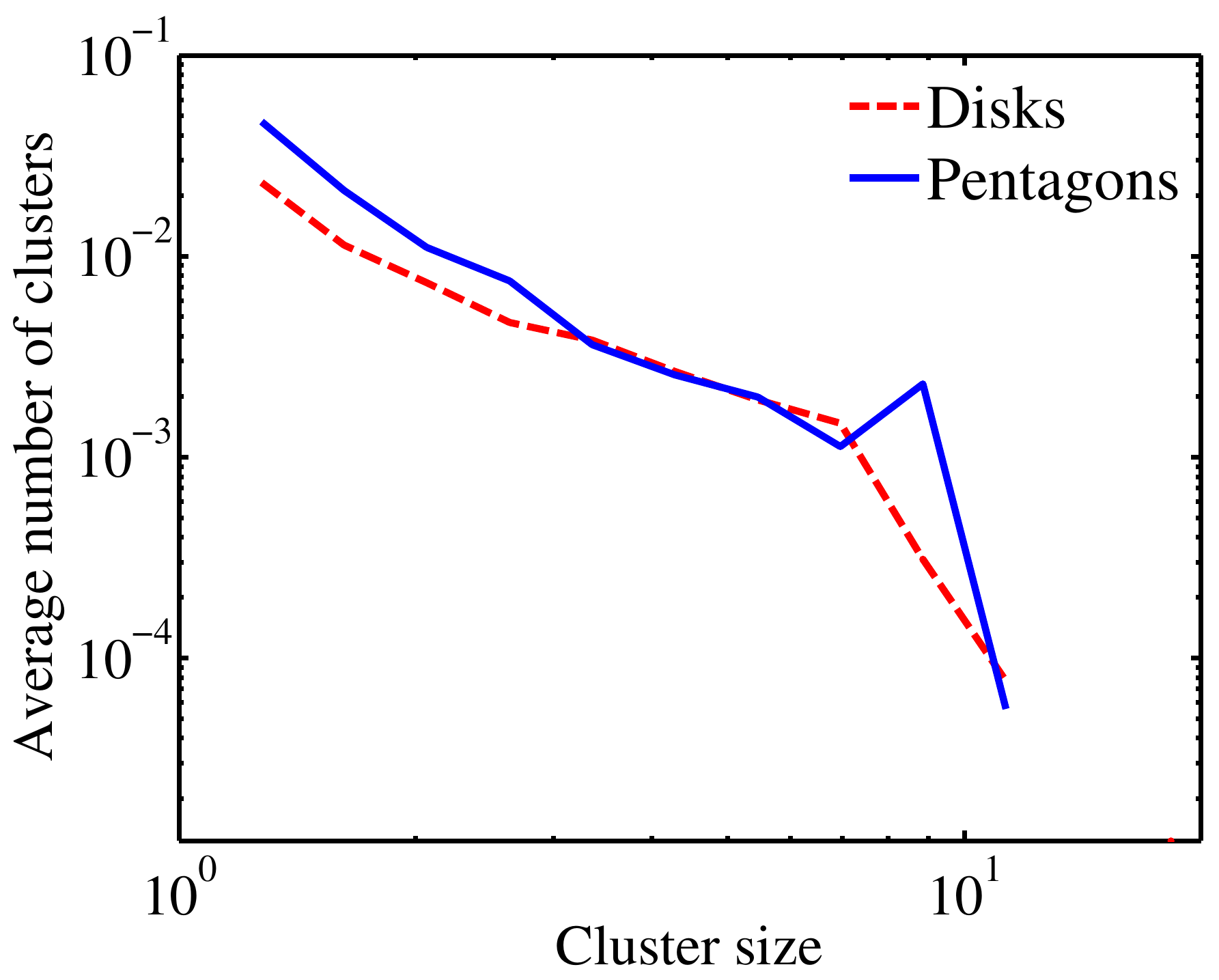}} \\
\subfigure[$F=1.0$]{\includegraphics[width = 1.6in]{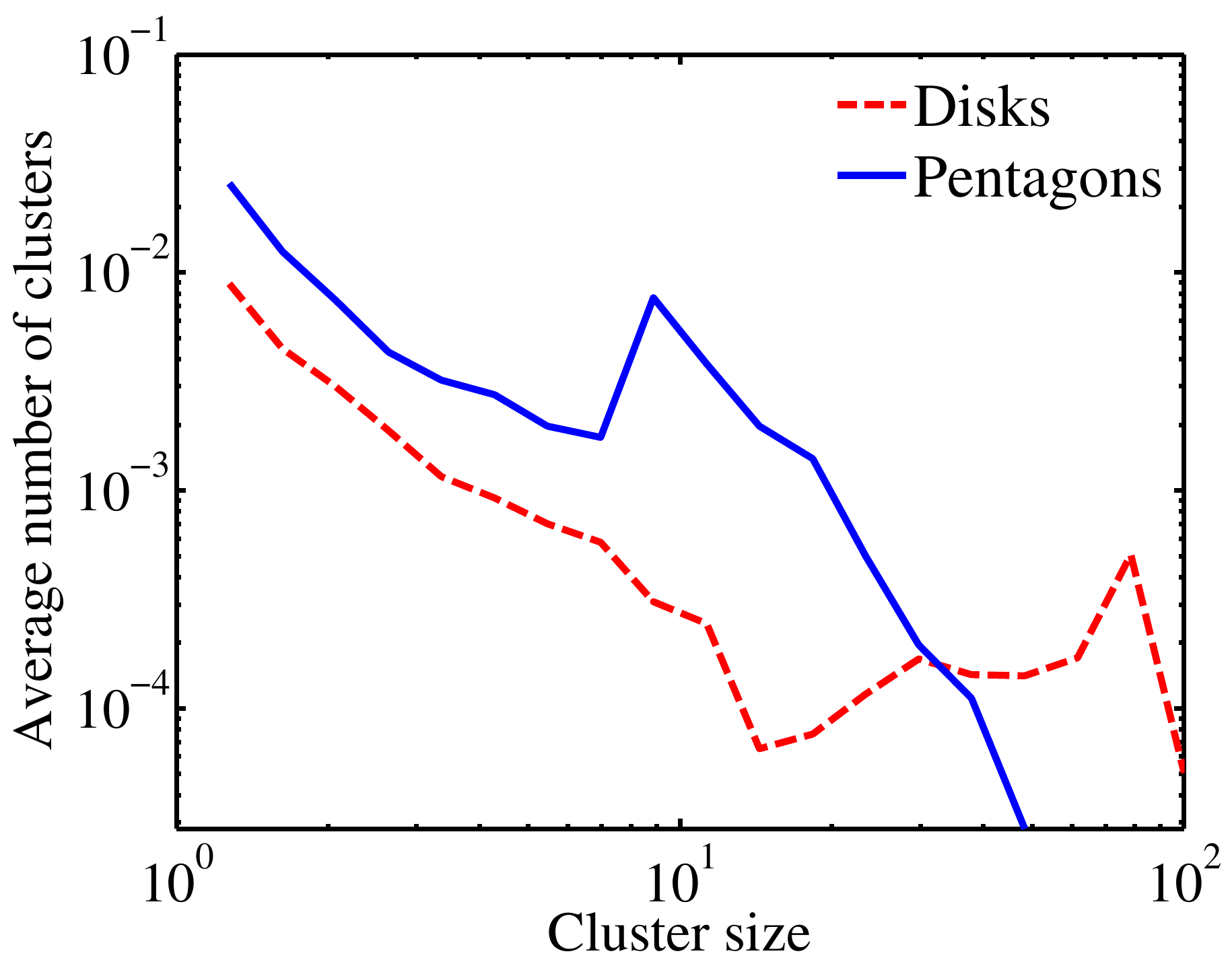}}
\subfigure[$F=0.5$]{\includegraphics[width = 1.6in]{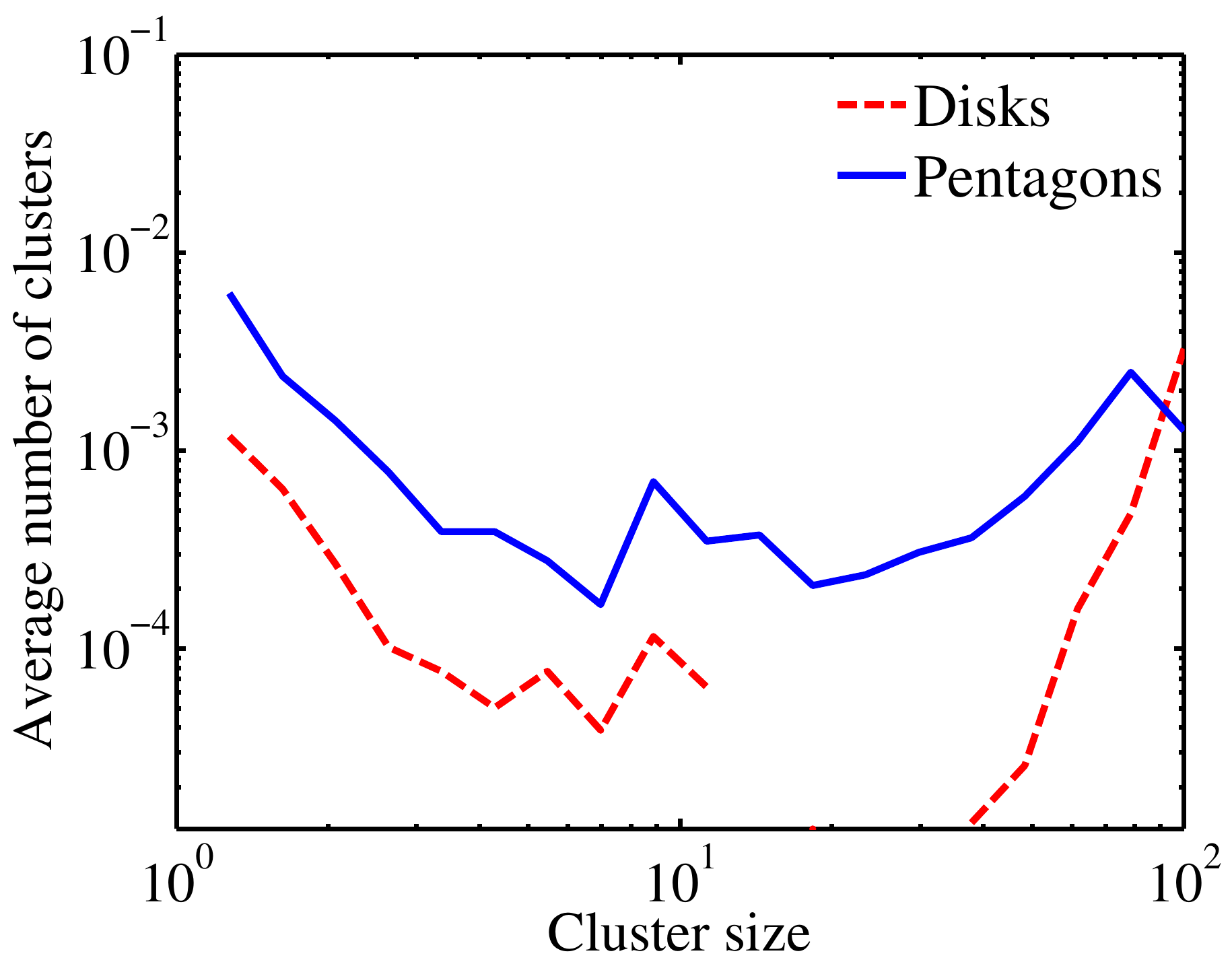}}

\caption{(Color online) Average number of components (clusters) per  particle for disks and pentagons in the tangential force network as a function of cluster size for 
various force thresholds, $F$. (a) $F=3$, (b) $F = 2.0$, (c) $F =  1.0$ and (d) $F= 0.5$ (bottom slice, low tapping).}

\label{fig:cluster_sizes}
\end{figure}

In order to highlight further differences between the systems composed of pentagons and disks, we focus on the tangential force networks, since these show clearer contrasts. We start by investigating the size of the  components at different force thresholds.  Figure~\ref{fig:cluster_sizes} shows the average number of components, per particle, as a function of component size for four different values of $F$. The shape of the curves changes significantly around $F = 2$.  Around this point, the  total number of  components, $\beta_0$, for both systems reaches its maximum [see Fig.~\ref{fig:disks_pents_b0}(b)]. For $F \geq 2$, the pentagon-base system tends to have a larger number of small clusters, consisting of less than approximately  $10$ particles (see Fig.~\ref{fig:cluster_sizes}), while the disk-based system contains a larger number of larger clusters. This suggests that for the disks, the contacts  with high forces tend to aggregate (in larger clusters).   These high force contacts are more scattered for the pentagons. 

As the value of $F$ is lowered below $2$, the   components start merging together.  Naturally,  the number of small clusters decreases  as they merge and create larger ones. However, there is a substantial difference between disks and pentagons. For $F=1$, pentagon packings do not contain any clusters composed of more than $\approx 50$ particles while for the disks a cluster  of size comparable to the number of particles
in the slice can be formed, viz. Fig.~\ref{fig:cluster_sizes}(c).  Again, the number of small clusters is larger for pentagons. Finally, for $F = 0.5$ the disks do not form any clusters of intermediate sizes. This is not the case for pentagons which exhibit clusters of all possible sizes.

We conclude that disks have a more heterogeneous structure of the tangential force network than pentagons. In disks packings, contacts with large tangential forces tend to appear spatially clustered and these clusters merge together for relatively larges value of $F$. On the other hand, pentagons form a large number of small spatially scattered clusters of high tangential forces. These scattered clusters grow and merge together  without spanning the system until $F$ reaches a very low value.

\begin{figure}
\centering
\subfigure[Normal forces with trivial loops.]{\includegraphics[width = 1.6in]{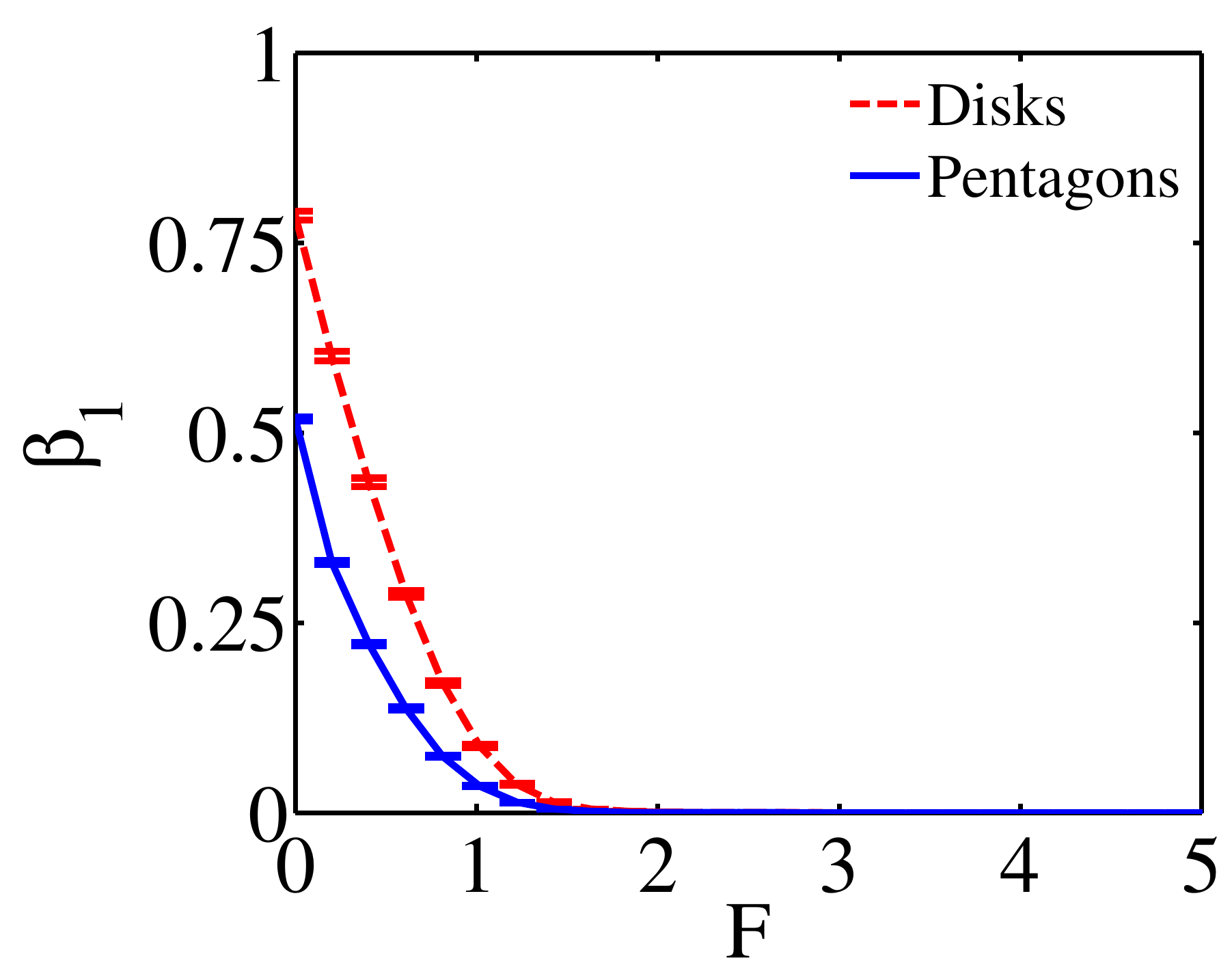}}
\subfigure[Tangential forces  with trivial loops.]{\includegraphics[width = 1.6in]{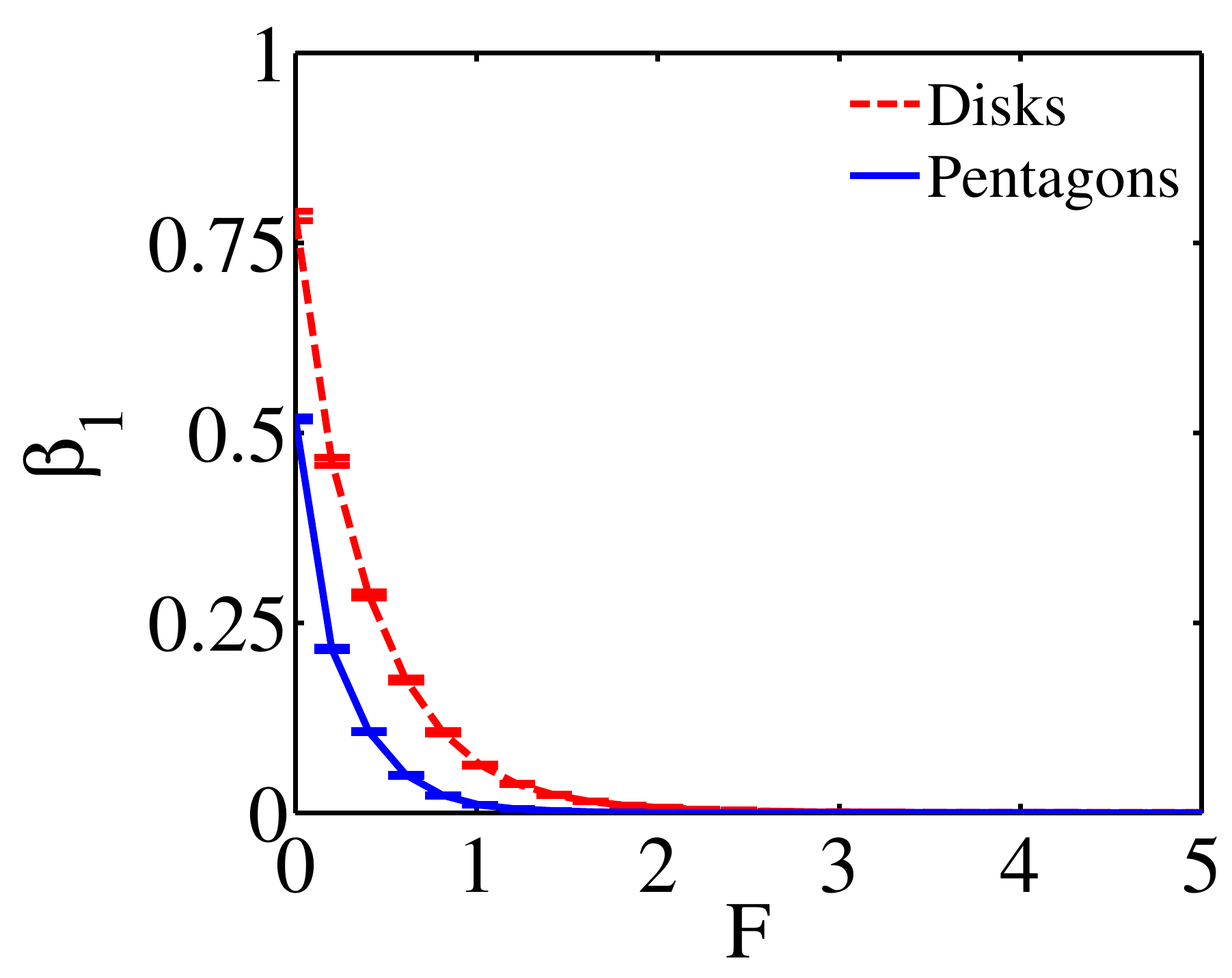}}
\subfigure[Normal forces  without trivial loops.]{\includegraphics[width = 1.6in]{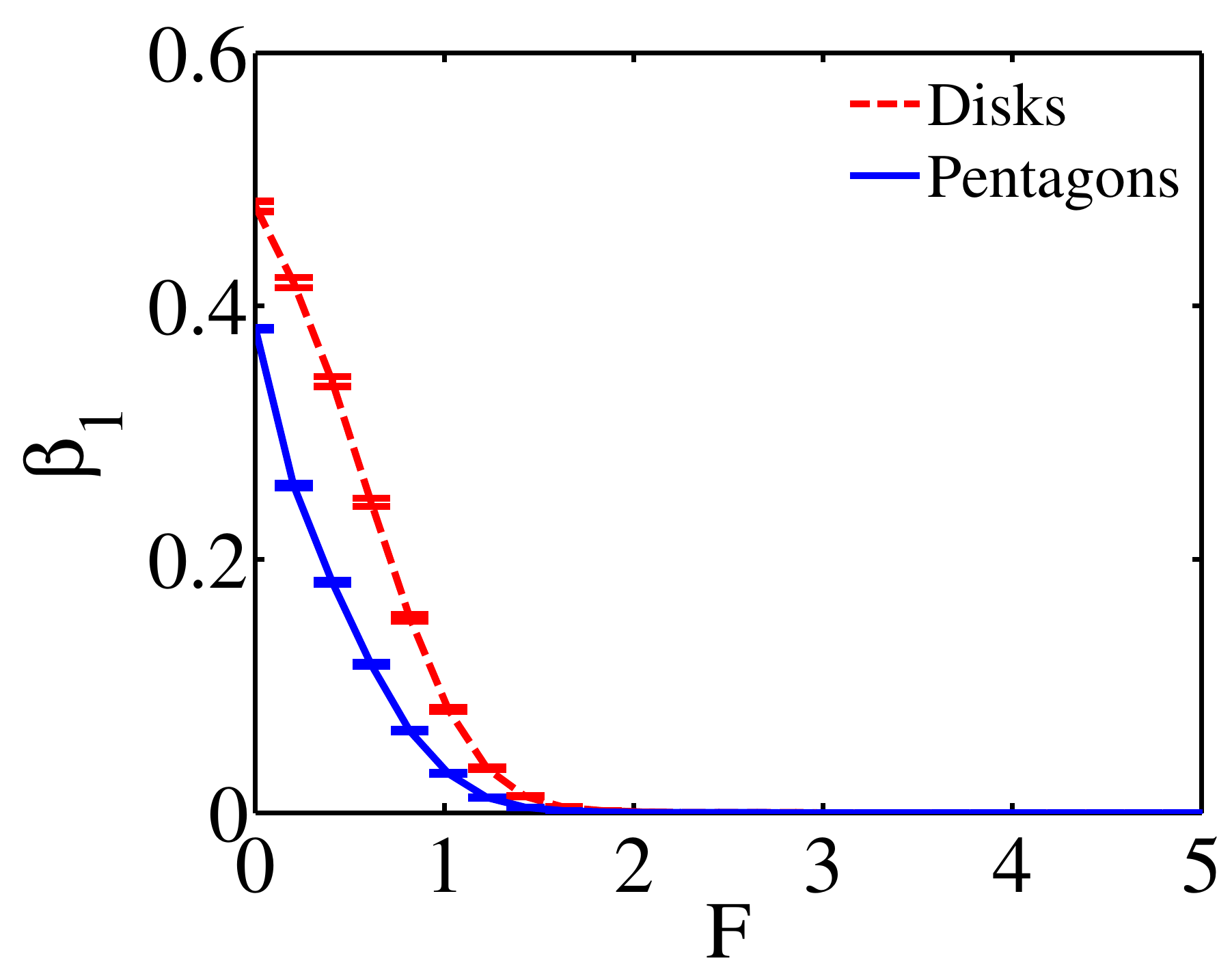}}
\subfigure[Tangential forces  without trivial loops.]{\includegraphics[width = 1.6in]{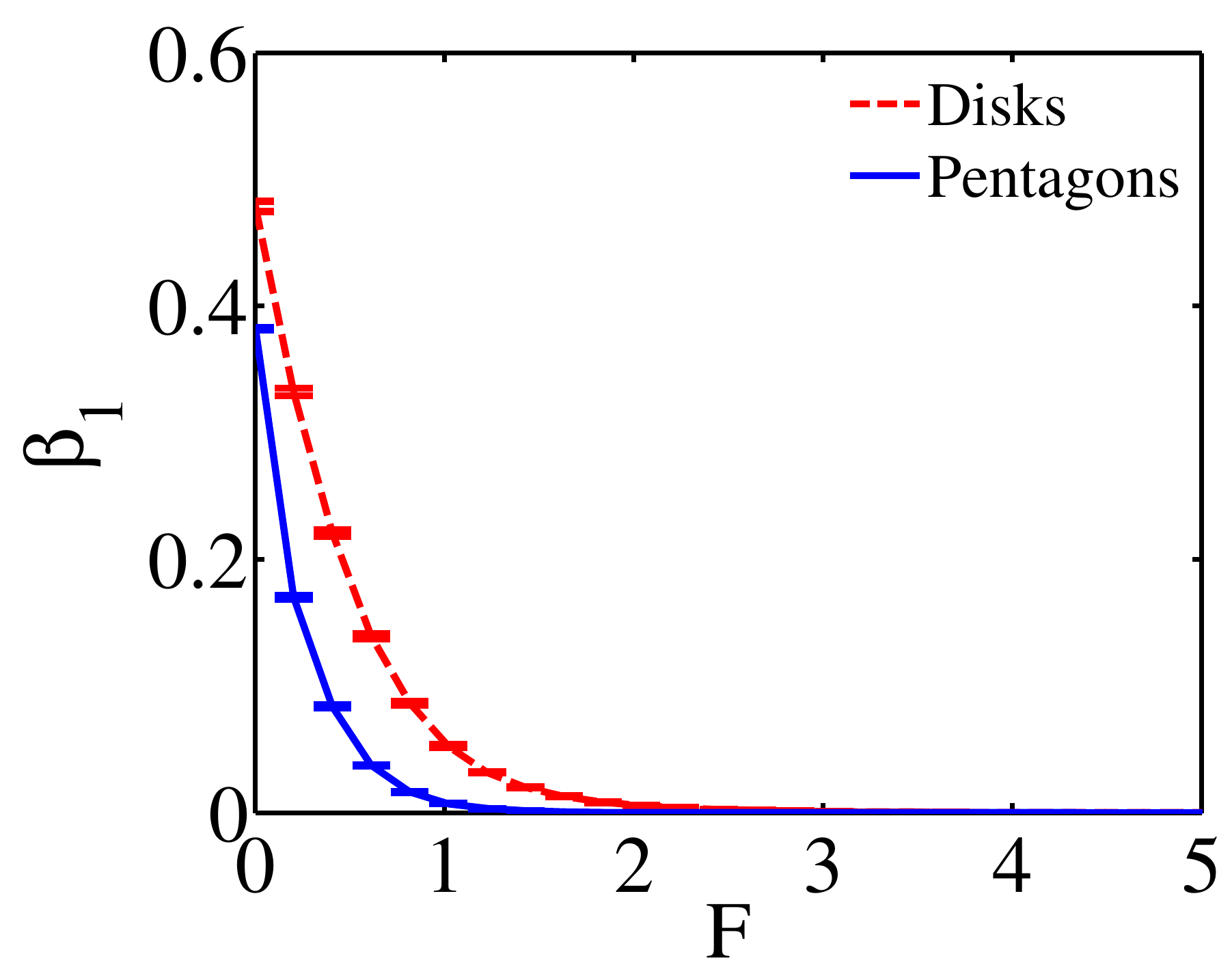}}
\caption{(Color online) $\beta_1$ as a function of the force threshold $F$ for disks and pentagons. The trivial loops are included in (a) and (b) and omitted in (c) and (d) (bottom slice, low tapping).
}
\label{fig:disks_pents_b1}
\end{figure}

Let us now consider the number of loops in the force networks described by the first Betti number, $\beta_1$. 
Figure~\ref{fig:disks_pents_b1}(a)~-~(b) show the number of all loops as a function of $F$.   
Figure~\ref{fig:disks_pents_b1}(c)~-~(d) presents similar data, but here the trivial loops formed by three particles in contact  are not included. In all the cases $\beta_1$ decays rather fast and there are only few loops for $F > 2$. This is as expected since for $F >2$ the clusters tend to be too small to contain a loop.

Disk packings contain a larger number of loops than pentagons  for both normal and tangential force. The number of trivial loops is significantly larger for the disks because of their  tendency to form crystalline regions. The larger number of loops observed in the tangential force network for disks even at $F>2$ is connected with the fact that these networks typically  contain larger  clusters that support a larger number of loops.  The number and sizes of the clusters in the normal force network are similar for both disks and pentagons, and thus,  the reason for a larger number of loops in the normal force network for the disks  is less clear.
We note that the differences in the loop structure of the force networks for disks and pentagons are consistent with the intuition that pentagons tend to form long arches that create large loops (loops of many edges) in the force network. Due to the large size of the loops their number tends to be smaller.
Finally, for $F=0$, the number of loops in both  normal and tangential  force networks equals  the number of loops in the contact network. We note that the number of loops present in the contact network for disks is consistent with an earlier study \cite{arevalo_pre13}.

\section{Conclusions}
\label{sec:conclusions}

We have studied the force network of static packings under gravity by averaging properties over a collection of realizations obtained by tapping. We focus on 
exploring significant differences of the force network when  particles of different shapes (disks versus pentagons) are considered, but have also explored difference between states obtained by using different tap intensities.

We have shown that the topology of the force network changes with depth.
To some degree, this can be inferred from the PDFs of the contact forces. 
However, the detailed analysis of  components sheds further light on this phenomena. 
The differences are not only in the average pressure, but also in the structural properties of the network.
Therefore, in this paper we analyze slices of the column to avoid artifacts due to the vertical heterogeneity  of the network.

We also compare realizations obtained using very different tap intensities that nevertheless have the same average packing fractions. 
These steady states have been  shown previously to present distinguishable mean stress \cite{pugnaloni_pre10} and distinguishable number of trivial loops in the contact network \cite{arevalo_pre13,ardanza2014topological}, albeit by using a different protocol and frictional walls.
For the present system, our results for the PDFs of the contact forces, force--force correlations, number of clusters and loops, cannot set apart in any significant way these equal-density states.   

Comparison of packings of particles of different shape have revealed that pentagons tend to exhibit a more  homogeneous tangential force network than disks. For the pentagon packings,  the edges in the force network,  corresponding to the strong contacts, are more spatially scattered  and less connected to each other.  In contrast, for the disks, the edges with smaller force value tend to be directly connected to the edges with larger values.   Hence,  the part of the tangential force network exceeding any given  force threshold, contains a larger number of small  components  for pentagons than for disks. In particular,  pentagon packings  contain many scattered `hot spots' in the tangential force network. On the other hand, for the disk packings, the part of the tangential force network exceeding larger values of $F$  consists of  smaller number of larger components.

These clear differences in the tangential force network do not have a matching effect in the normal force network. Differences in the normal force network between disks and pentagons are hard to detect. Interestingly, shorter force chains, consistent with smaller clusters have been observed for pentagons in comparison with disks in experiments involving shearing (see~\cite{behringer_2001}, Fig. $26$). Notice however that in this experimental study only normal forces are considered.

One significant difference between disk and pentagon packings is the number of loops. 
The existence of a larger number of trivial loops (formed by three grains in mutual contact) in the case of disks is not particularly surprising. However, pentagons also display a smaller number of non-trivial loops. This suggests that pentagons tend to form rather large loops, particularly when tangential force networks are considered.

While the approach used in the present paper has uncovered a number of properties of force networks, it should be pointed out that counting
of the number of (connected) components and loops as a function of force threshold does not contain the whole picture: knowing Betti numbers
does not tell us how the connections of the features (components, loops) evolve as force threshold is varried.  For this purpose, we need to 
consider additional measures, based on persistent homology.  As we will see in~\cite{paper2}, this approach uncovers significant additional
features of force networks, and in particular allows to quantify the differences between the systems exposed to different tap intensities that
lead to the same packing fraction.   In addition, persistent homology will allow us to describe and even measure the differences in the force networks 
from one realization/tap to the next.   

In the present work, we have focused on the influence of gravitational compaction, tapping intensities leading to similar packing fraction, 
and particle shape on the properties of the force networks. However, we have not studied the effects of boundary conditions or system size, and 
have not systematically explored the general influence of the tapping intensity on the properties of force networks.  These directions of research will 
guide our further investigations. 

\begin{acknowledgments}
KM and MK were partially supported by NSF grants No. DMS-0915019, 1125174, 1248071, and contracts from AFOSR and DARPA.  LK acknowledges support by the NSF grants No. DMS-0835611 and DMS-1521717.  
\end{acknowledgments}

\bibliographystyle{apsrev}

\end{document}